\documentclass{raa}            

\usepackage{graphicx,times}             
\usepackage{natbib}
\usepackage{amssymb,amsmath}
\bibpunct{(}{)}{;}{a}{}{,}

\usepackage[a4paper=true,pagebackref=true]{hyperref}
\hypersetup{colorlinks = true, linkcolor = green, anchorcolor = red, citecolor = blue, filecolor = red, pagecolor = red, urlcolor = red}

\begin{document}

   \title{Site testing campaign for the Large Optical/infrared Telescope of China: Overview}
   

   \volnopage{Vol.0 (20xx) No.0, 000--000}      
   \setcounter{page}{1}          

   \author{Lu Feng\thanks{E-mail: jacobfeng@bao.ac.cn}\inst{1},
Jin-Xin Hao\inst{1},
Zi-Huang Cao\inst{1,7},
Jin-Min Bai\inst{2},
Ji Yang\inst{6},
Xu Zhou\inst{1},
Yong-Qiang Yao\inst{1},
Jin-Liang Hou\inst{4},
Yong-Heng Zhao\inst{1},
Yu Liu\inst{2},
Teng-Fei Song\inst{2},
Li-Yong Liu \inst{1},
Jia Yin\inst{1,7},
Hua-Lin Chen\inst{5},
Chong Pei\inst{5},
Ali Esamdin\inst{3},
Lu Ma\inst{3},
Chun-Hai Bai\inst{3},
Peng Wei\inst{3},
Jing Xu\inst{3},
Guang-Xin Pu\inst{3},
Guo-Jie Feng\inst{3},
Xuan Zhang\inst{3},
Liang Ming\inst{3},
Abudusaimaitijiang Yisikandee\inst{3},
Jian-Rong Shi\inst{1},
Jian Li\inst{1},
Yuan Tian\inst{1},
Zheng Wang\inst{1,7},
Xia Wang\inst{1},
Xiao-Jun Jiang\inst{1},
Jian-Feng Wang\inst{1},
Jian-Feng Tian\inst{1},
Yan-Jie Xue\inst{1},
Jian-Sheng Chen\inst{1},
Jing-Yao Hu\inst{1},
Zhi-Xia Shen\inst{1},
Yun-Ying Jiang\inst{1}
}

   \institute{National Astronomical Observatories, Chinese Academy of Sciences,
             Beijing 100012, China; {\it jacobfeng@bao.ac.cn}\\
        \and
             Yunnan Observatories, Chinese Academy of Sciences, Yang Fang Wang 396, Kunming, Yunnan, China\\
        \and
             Xinjiang Astronomical Observatory, Chinese Academy of Sciences, 150 Science 1-Street, Urumqi, Xinjiang, China\
        \and
            Shanghai Astronomical Observatory, Chinese Academy of Sciences, Nan Dan Lu 80, Shanghai, China\\
        \and
            National Astronomical Observatories\/Nanjing Institute of Astronomical Optics \& Technology, \\Chinese Academy of Sciences, Nanjing 210042, China
        \and
            Purple Mountain Observatory, Chinese Academy of Sciences, Ban Cang Jie 188, Nanjing, China
        \and
            University of Chinese Academy of Sciences, Beijing, China\\
\vs\no
   {\small Received~~20xx month day; accepted~~20xx~~month day}}

\abstract{ The Large Optical/infrared Telescope (LOT) is a ground-based 12m diameter optical/infrared telescope which is proposed to be built in the western part of China in the next decade. Based on satellite remote sensing data, along with geographical, logistical and political considerations, three candidate sites were chosen for ground-based astronomical performance monitoring. These sites include: Ali in Tibet, Daocheng in Sichuan, and Muztagh Ata in Xinjiang. Up until now, all three sites have continuously collected data for two years. In this paper, we will introduce this site testing campaign, and present its monitoring results obtained during the period between March 2017 and March 2019.
}

   \authorrunning{L. Feng et al}            
   \titlerunning{Site testing for LOT: overview}  

   \maketitle

%
%
\section{Introduction}           
\label{sec:background}
Astronomy in China is growing rapidly in recent years. After a community-wide survey in 2015, a new project, the Large Optical/infrared Telescope (LOT) was elected to be built in the following decade. The telescope is a general purposed telescope with a unique 4-mirrors design \cite{Su2016}. Its primary mirror is consisted of 84 hexagonal segments with a diagonal length of 1.44m for each of these segments. The telescope will provide prime focus, Cassegrain, Nasmyth and Coude foci for mounting scientific instruments. Three first light instruments were planned to be installed at the telescope's Nasmyth focus: a Broad band Medium Resolution Spectrograph (BMRS), a Wide-field Imaging SpEctrograph (WISE), and a High Resolution Spectrograph (HiReS). All three first light instruments are seeing limited. Wavelength coverages are the same for all three instruments, which is 340nm to 1000nm \cite{Cui2018}. Adaptive optics and diffraction-limited infrared instruments were planned as future upgrades of the telescope. In order to maximize the performance of the telescope and its instrument suite, it is important to find a suitable site that is excellent both from its astronomical performance as well as operation and maintenance perspectives.

Early efforts have been made by the community and have been reported in papers such as \cite{Liu2012, Yao2013, Yao2015, Liu2015, Wang2015, Qian2015, Wu2016, Liu2016}. However, due to differences in data definitions, data acquisition instruments, and data processing methods, etc., results from these various site testing programs are not directly comparable and cannot help in reaching a consensus conclusion. Therefore, immediately after the go-ahead decision of LOT project in June, 2016, a decision to start a long-term site testing campaign was made, and related works soon started in late 2016.

Considering available site testing instruments at that moment, priority was given on monitoring sites' astronomical performances in visible band in the early phase of the campaign. Parameters such as cloud cover, optical seeing, night sky brightness, meteorological parameters, were required to be measured from the start. Relevant equipment were purchased, calibrated and installed within a short time frame, and by the beginning of 2017, all three candidate sites have already begun routine site monitoring activities continuously. Precipitable Water Vapor (PWV) was added into the monitoring parameter list in early 2018. Other infrared or Adaptive Optics (AO, which still has no decent correcting performance for shorter wavelength than the infrared) related site performances, such as vertical turbulence profile, wind vector profile, sodium layer characteristics, etc., are planned to be measured in near future.

This paper is a general introduction of the results from this two years sites monitoring campaign. Several other papers \cite{Cao2019a, Cao2019b, Liu2019, Song2019, Xu2019a, Xu2019b, Wang2019} detailing the site testing campaign's different aspects will also be published in the same special issue. We would like to refer these papers to our readers for further details.

In section \ref{sec:selection}, we will first briefly discuss how these three long-term monitoring sites were chosen. In section \ref{sec:site} we will describe the technical aspects of these candidate sites, especially, parameters that have been monitored, relevant site testing instruments and their setup at each site, as well as data processing methods. In section \ref{sec:results}, we will show sites' monitored results of these two years (March 2017 - March 2019). We will discuss what have been learned during the first year's operation in section \ref{sec:discussion}.

\section{The selection of the three long-term monitoring sites} \label{sec:selection}

During 1980s-1990s, survey results for astronomical sites have already been in favor of the western part of China. Several sites have been selected and established in this period as a result, such as: Delingha site in Qinghai (operation started since 1983), Nanshan site in Xinjiang (1991), Gaomeigu site in Yunnan (1995). Altitudes of these sites are among 2080-3200m above sea level, much higher than previous sites in China which are usually located near sea level. The application of these sites result in a significant improvement of the performance of astronomical telescopes in China. In 2003, a dedicated site selection team was organized within NAOC, aimed particularly for investigating sites for the next-generation telescopes in optical and other wavelengths. Large scales studies started from meteorological stations' data and cumulative cloud maps from satellite data. With these analysis, and on-site inspections, the survey was focused to three sites for long-term monitoring during which include Karasu in Pamir Plateau, Oma and Gar sites in western Tibet. In 2010, a solar site survey for the next generation large solar telescope was discussed. Therefore, a site survey aiming for the project was initiated and this time, Ali of Tibet, Ganzi (Daocheng) in Sichuan, and Deqing in Yunnan, were inspected particularly for their day-time performance.

\begin{figure}
\includegraphics[width=\columnwidth]{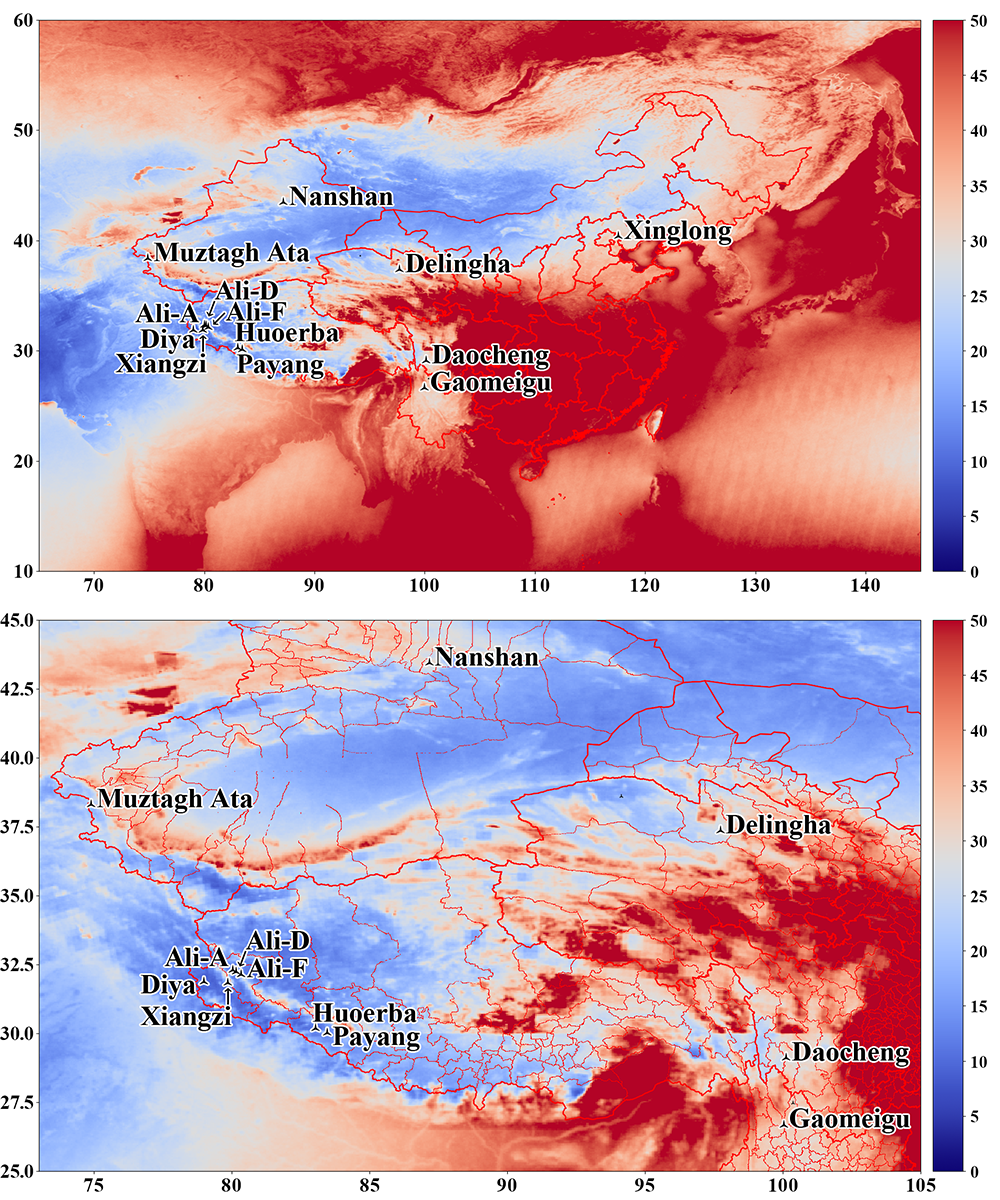}
\caption{Annually averaged probability of cloudness at night estimated from MODIS data set. (Upper) Nation-wide estimation with probability color mapped from 0\% to 50\%. (Lower) Zoomed-in view of the estimation for the western part of China with probability color mapped from 0\% to 50\%. X and Y axes are the longitude and latitude respectively. Color from blue to red represents the possibility of a cloud night from good to bad.}
\label{fig:remote sensing 1}
\end{figure}

A unified site testing campaign was initiated in 2016 by the LOT site testing work group. The first step conducted by the team was to utilize long term remote sensing data from satellites to analyze nation-wide large scale cloud coverage. Two independent teams analyzed the cloud coverage, and each team utilized different data sources and adopted independent analyzing methods \cite{Cao2019b}. The MODIS satellite data set provides two instantaneous images every night at local time of 22:30 and 01:30 from 2003 to 2015, with a pixel scale of $5$km$\times$$5$km. The GMS+NOAA satellite data set provides images for every hour with a pixel scale of $36$km$\times$$36$km. The GMS+NOAA data set has record from 1996 to 2003, much dated comparing to the MODIS data. Pixel scale for both data set are not small enough for pin-point optimum locations for monitoring sites. However, cloud estimation even with this large scale can indicate optimum regions where potential sites might be found. Besides, by examining the long term coverage of these satellite data set, development trend of cloud at interested sites could also be observed which could be helpful for predicting cloud coverage in these regions for near future. Considering GMS+NOAA data covers 1996-2003 with coarser spatial resolution while MODIS covers a more recent span of 2003-2015 with finer resolution, the two data sets are complementary for our purpose. In paper \cite{Cao2019b}, Cao et al. discussed more in detail on the analysis methods, cross-comparison  results between these two data sources and data from ground based instruments, the evolution of cloud in China based on the data and the possible driving forces that contribute to this evolution. We kindly refer our readers to their paper for details. Here in this paper, we only excerpt part of their results as shown in figure \ref{fig:remote sensing 1}.

Figure \ref{fig:remote sensing 1} shows the annually averaged probability of cloudness at night estimated from 2003-2015 MODIS data. Generally speaking, a night at a pixel is justified as cloudy night if that pixel in both MODIS images at 22:30 and 01:30 are 100\% covered with clouds. From the upper plot of the figure which shows the nation-wide cloud estimation, the western part of China has significant less clouds. In the zoomed-in plot, color coded with a smaller range (from 0-50\%), there are several bluer strips which indicate even lesser cloud possibilities along the western border of China. Theses regions are where Ali and Pamir plateau (Muztagh Ata, may also be refered as Muztagh Ata) are located. In this plot, there is also a blue dot surrounded by red region in the southeast where Daocheng is located. 

\begin{table}
\centering
\begin{tabular}{cccc}
\hline
 & & GMS+NOAA (\%) & MODIS(\%)\\
Region&Epoch & 1996$\sim$2003 & 2003$\sim$2015\\
\hline\hline
Ali& & 74.5 & 86.8 \\
Daocheng& & 71.7 & 71.1 \\
Muztagh Ata& & 58.7 & 77.8 \\
\hline
Xinglong& & 64.4 & 67.9\\
Delingha& & 65.6 & 74.1\\
Gaomeigu& & 64 & 66.9\\
Nanshan& & 60.3 & 70.7\\
\hline
\end{tabular}
\caption{Estimation of annual percentages of clear nights for different regions from GMS+NOAA and MODIS data sets (1996-2015).  Four existing sites are also listed in the second part of the table.
}
\label{tab:GMS}
\end{table}

In table \ref{tab:GMS} we listed the estimated fractions of annual percentage of clear nights comparing to the total amount of nights for several sites from year 1996 to 2015 with each data set. Existing sites like Xinglong and Delingha are also included for comparison. One interesting thing to note in this table is the significant increase of clear nights at Muztagh Ata as well as Delingha and Nanshan during 2003$\sim$2015 comparing to 1996$\sim$2003. Cao et al. in their paper explained several climatic factors that may lead to this behavior. 

Estimation of cloud coverage from satellite data has advantage in spatial coverage, and it is helpful for identifying optimum regions within China. However, the spatial resolutions of this method are not sufficient to identify the best location within its kilometers-wide single pixel. The different definitions of cloudness between satellite and ground based observation would also lead to differences in the judgement of cloud. The cadence of the satellite data is also in the scale of hours which is not enough to catch fast variations of parameters like wind, temperature, low level clouds, etc. Lacking of other instruments on satellite to measure regular astronomical climate performances such as wind, relative humidity, temperature at ground, as well as seeing, transparency, etc., also limit the applicability to use this single method to evaluate sites' overall performance. Therefore, long-term ground-based astronomical climate monitoring campaign is required. Combining with other considerations such as: geographical information, traffic/logistics, meteorological parameters and previous site testing results, three sites (listed in Table 2) were chosen at last for long term monitoring for the site testing campaign.

\section{The candidate sites, testing instruments and setup} \label{sec:site}
\begin{figure}
\includegraphics[width=\columnwidth]{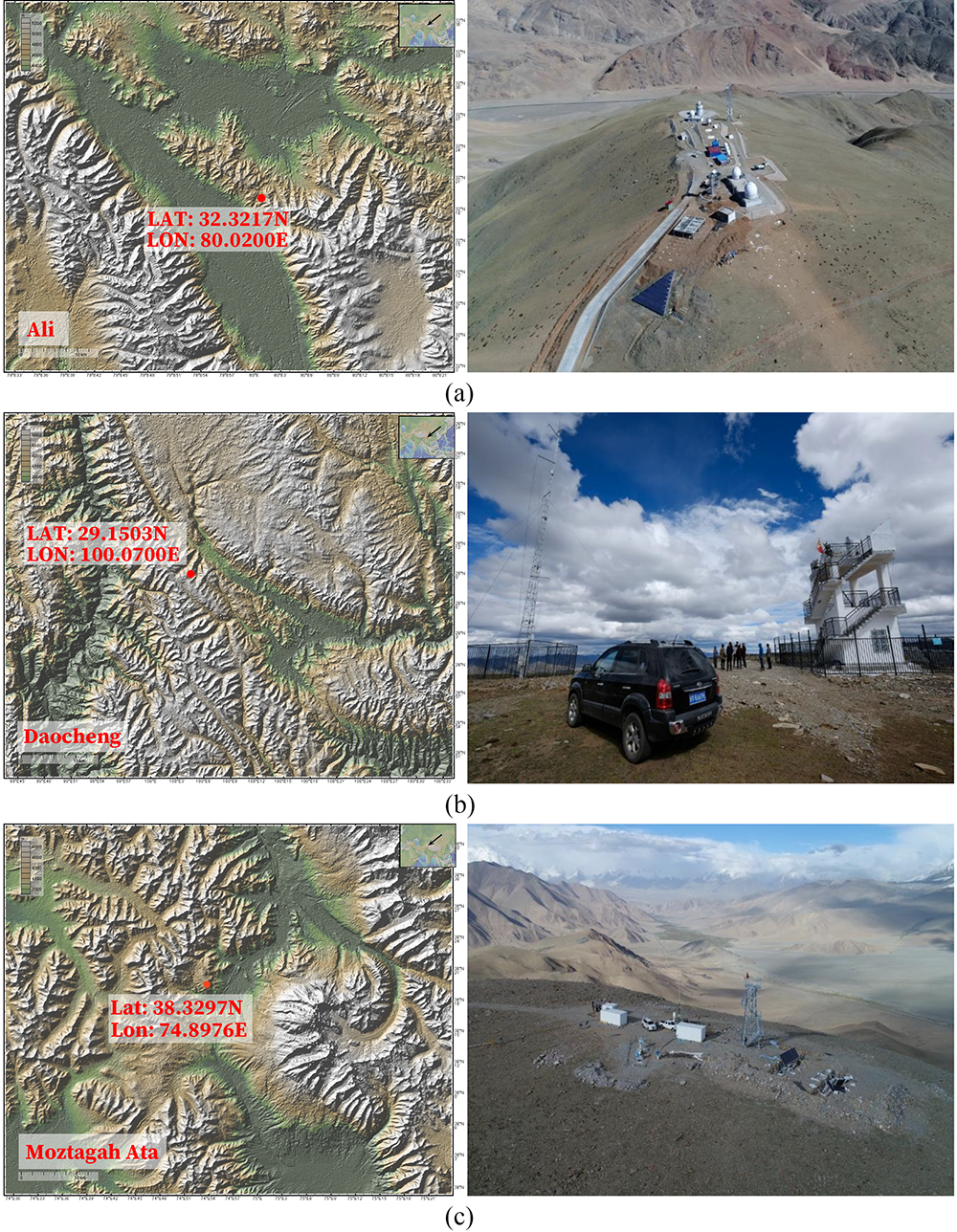}
\caption{Topographic maps (left) and photographies (right) of the candidate sites. (a) Ali, (b) Daocheng, (c) Muztagh Ata. All topographic maps have the same distance scale and cover 70kmx80km area. Red spot in each topographic map shows the location of these sites. Colors from green to white indicate relative heights from low to high within each map.}
\label{fig:topo site}
\end{figure}

Table \ref{tab:location} lists the three sites that were chosen and their coordinates. Figure \ref{fig:topo site} shows their local topographical maps and photos. Both Ali and Daocheng sites have already been constructed and being monitored \cite{Yao2014,Yin2015,Liu2015,Wang2015,Yao2012,Liu2012,Yao2013,Wu2016,Liu2016} for a relatively long time. Preparation works for the Muztagh Ata site only began after the initiation of LOT project in late 2016. Due to the low temperature at Muztagh Ata during that winter, construction works especially concrete works were not finished until March 2017. Figure \ref{fig:topo site} shows the bird-view of these sites. 

\begin{table}
\centering
\begin{tabular}{p{0.23\columnwidth}cccp{0.14\columnwidth}}
\hline
Site's name & Alitutde &  Longitude & Latitude & Established\\
\hline\hline
Ali & 5100 & 80.046E & 32.306N & 2012\\
Daocheng & 4750 & 100.109E & 29.107N & 2016 \\
Muztagh Ata & 4526 & 74.897E & 38.330N & 2017\\
\hline
\end{tabular}
\caption{Altitude and location of the monitoring sites}
\label{tab:location}
\end{table}

The Ali site is located in southwest of Tibet province at the tip of a mountain range. It has an altitude of 5100m. The closest town that could provide accommodation and be utilized as headquarter is Shi-Quan-He. The town is about 30km away north from the site with an average altitude of 4300m, and it has a population of 20,000. Ali Kun-Sha airport is on the south side of the site. Paved road linking both the airport and the town passes by the site. A branching path from this road to the site has been built. The current location of the site is called point A. Facilities, electricity and network have already been provided on site. Several small telescopes with aperture diameters ranging from 0.3m to 1.0m are also there in operation. Another two locations (point B and C, with altitudes of 5160m and 5380m respectively) on the same mountain range in the downwind direction (southeast) are still under development by the time of this paper. 

The Daocheng site is located between Ganze county and Daocheng county in Sichuan province. The altitude of the site is 4750m. The closest town is Daocheng, $\sim$27km away from the site, with an altitude of 3750m. The Ganze county is $\sim$65km away, at a much lower altitude of 2750m. The population of Daocheng is 7900. Although electricity and cable network have not reached the site yet, generator and solar panel have already been built and could provide electricity to all equipments while data transfer still requires personnels to download at the site and upload to server after their returns to town. A concrete observation tower for site's performance monitoring has already been built and all equipments on its top deck are in operation. 

The Muztagh Ata site is a newly selected site, with no facility or equipment set up previously. The site is located on Pamir Plateau and is 11km away from the 7546m-high Muztagh Ata Mountain which is famous among mountain climbers. The site's altitude is 4526m, and it is about 220km away from the city of Kashgar. The closest town is Bulunkouxiang at an altitude of $\sim$3300m, 50km away in the north.  

It was decided at the beginning that during the site testing campaign, all sites should use the same or at least identical equipments so comparable measurements could be obtained and be processed with the same data processing method by an independent data analysis team. The top priority is to have comparable visible band related parameters as well as parameters involving the design and operation of the telescope for its early phase while additional equipments could be added into the list to measure infrared related parameters in near future. The following part listed the parameters being monitored and instruments used for these measurements.

\begin{enumerate}
\item \textbf{Temperature, atmospheric pressure, relative humidity, wind direction and speed}: These are climatic parameters that represent local environment's characteristic. They are important parameters for the design and construction of the facility. The reasons to monitor these parameters from astronomical perspective are the following. Environmental temperature variation between night and day will have impact on dome seeing when the dome is opened at night. Even when dome temperature is settled, because large portion of the noise in infrared observation comes from thermal noise of the telescope structure which is at the temperature of the environment, it would be better if the environmental temperature of the site is low and stable. Temperature, atmospheric pressure and relative humidity combined could cause dew to condense on the surface of mirror interfering observation. Wind near ground will induce shake and vibration to the structure of the dome, and most importantly, to the surface of the mirror, so it is important to know the strength and direction of the wind. These parameters are measured continuously with commercial automated weather station \cite{huayun}. The weather station produces 1 value for each of these parameters every minute. Measurements were stored locally, copied out and transmitted manually to a central server by local crew. One of these weather stations was installed at each of the sites.
\item \textbf{Cloudness}: Cloud in the line of sight of the telescope will interfere observation. It would reduce the flux arriving at the telescope to certain degree depending on the amount of cloud during long exposure, or even making the observation impossible to proceed. Therefore, it is important to know the typical statistics of cloud coverage within certain angles from zenith at night for all year. As for the campaign, cloudness is monitored by All Sky CAmera (ASCA) which is designed and fabricated by National Astronomical Observatories, Chinese Academy of Sciences (NAOC) \cite{Wang2019}. The same type of ASCAs were installed at all sites. The camera has two parts, a Sigma 4.5mm fish eye lens and a Canon 700D camera providing a $180^\circ\times180^\circ$ field of view. The ASCA has no filter. Its sampling frequency in day time is 1 picture for every 20 minutes with exposure time of 1/3200 of a second. At night time, sampling frequency is increased to 1 picture for every 5 minutes with exposure time adjusted to a value between 15 and 30 seconds depending on lunar phase. All raw images were stored and transmitted to a central server by local crew everyday. The independent data analyzing team used a similar manual cloud identification method as described in Thirty Meter Telescope's site testing campaign paper \cite{Skidmore2008}. Same standard is applied to all ASCA images for all the candidate sites to ensure comparability. More details on the data reduction method can be found in \cite{Cao2019a}.
\item \textbf{Sky background in V band}: Sky background is one of the main noise sources for observation in optical wavelength. To evaluate sites' V band sky brightness, commercial Sky Background Meters (SBM), Unihedron SQM-LE \cite{Unihedron2016} was used. The SBM is mainly a high sensitivity photo-diode with a field of view of $\sim$20$^\circ$ performing photometry in the zenith direction every minute at night time. A visible band filter is installed inside the instrument and the measurement is automatically converted into Johnson V band. SBMs were set up at all sites, generally next to the ASCAs.
\item \textbf{Night seeing at 500 nm}: Local seeing reflect the total integral strength of atmospheric turbulence in the line of sight of telescope. It will affect the image quality of the telescope if without the correction from adaptive optics. To be able to compare between different sites, seeing is defined as the best angular resolution in the zenith direction with 0 second exposure time that a site can achieve. An instrument called Differential Image Motion Monitor (DIMM) as described in \cite{Sarazin1990} is commonly used to measure this value. In the site testing campaign, three models of DIMMs were deployed at the candidate sites as shown in table \ref{tab:DIMMs}. All sites were equipped with at least two DIMMs, one for continuous measurement, the other one acting as a backup if the first one fails. An extra DIMM was first calibrated for two weeks with a DIMM that had been in operation for years at Xinglong observatory. It was then installed side by side with all other DIMMs for cross calibration for at least two weeks for each site to ensure all measurements are comparable. 
\item \textbf{Precipitable water vapor}: Water vapor in the atmosphere acts as an absorptive medium for the infrared band, thus it is necessary to monitor the content of the water vapor above the sites. Column density of Precipitable Water Vapor (PWV) is commonly used for reflecting contents of atmospheric water vapor. To monitor it, we used a commercial Low Humdity And Temperature Profiling Radiometers (LHATPRO) from RPG \cite{rpg}. The radiometer measures vertical profiles of atmospheric temperature and humidity and based on these measurements, estimates vertical integrated water vapor for every second. The advantage of LHATPRO is that it could be used for night measurement because it operates in radio band than other PWV monitor instruments which rely on the measurement of solar extinction. However, the cost of the LHATPRO is not affordable to equip this instrument for all three sites. Because seasonal PWV variation monitoring has a higher priority than daily variation, only one LHATPRO was deployed recently and was shifted around the three sites for short term periodic monitoring.   
\end{enumerate}

\begin{figure}
\includegraphics[width=\columnwidth]{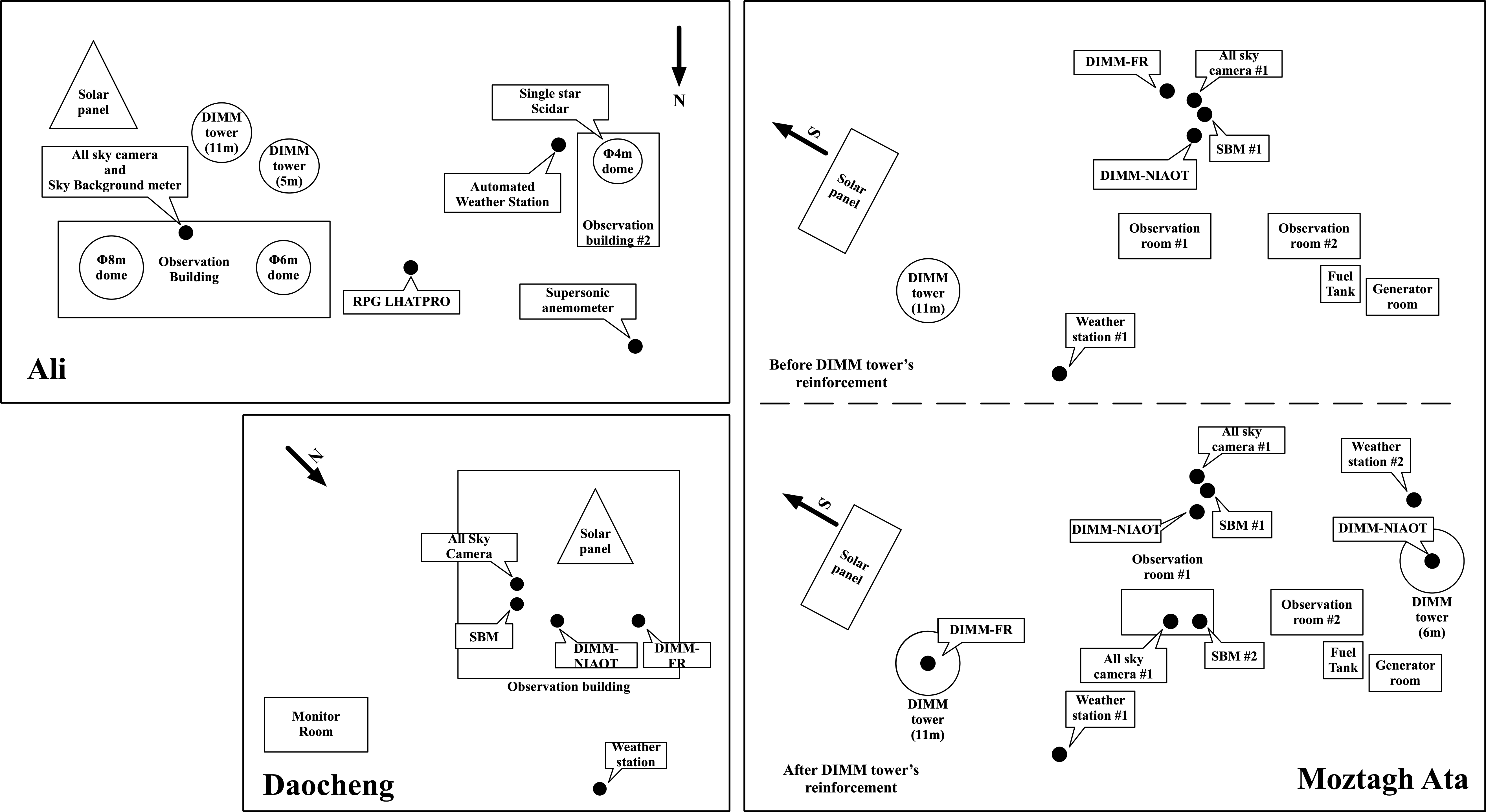}
\caption{Instrument setups at candidate sites. Ali site (Upper left) Daocheng site (Lower left) and Muztagh Ata (Right)}
\label{fig:instrument setup}
\end{figure}

\begin{table*}
\centering
\begin{tabular}{p{0.25\textwidth}p{0.2\textwidth}p{0.2\textwidth}p{0.2\textwidth}}
\hline
Model name & Ali DIMM & NIAOT DIMM & French DIMM\\
\hline\hline
Availability & custom & custom & \cite{alcor}\\
Telescope &MEADE LX200&GSO RC 8"& \\
Focal ratio &10&8&8\\
Focal length (mm) &2500&1600&2400\\
Subapertures & 2 & 4 & 2\\
Subaperture diameters (mm) & 50 & 50 & 50\\
Subaperture distance (mm) & 149 & 200 & 240\\
Camera & Basler aca2040 & Lumenera SKYnyx2-0M & DMK 33GX174 \\
Exposure method & 10ms & 10ms & between 0.5ms and 1000ms, adjusted automatically\\
Wavelength (nm) & 500 & 500 & 550 \\
Output frequency  & 1 seeing value for every minute & 1 seeing value for every minute & 1 seeing value for every 1000 images\\
Scaling and conversion& convert to zenith, no exposure time scaling & convert to zenith, no exposure time scaling & convert to zenith, no exposure time scaling, no wavelength scaling\\
Sites equipped & Ali & Daocheng and Muztagh Ata & Ali, Daocheng and Muztagh Ata\\
\hline
\end{tabular}
\caption{Information of DIMMs that were equipped at candidate sites}
\label{tab:DIMMs}
\end{table*}

Figure \ref{fig:instrument setup} (upper left) showed the instruments setup at Ali site. Initially an 11m tower was built for DIMM measurement. However, it was soon found out that the stability of the tower can not meet the requirement for DIMM measurement if the DIMM were installed on the top of the tower \cite{Cao2019a}. The structure of the tower was reinforced soon after. However, a few months of data for 11m tower was missing because of this. Meanwhile, another Ali DIMM was kept monitoring on the 3m height ceiling of the observation building. ASCA and SBM were also installed on the ceiling. A 5m tower was built in May 2017, and the DIMM that was previously installed on the 3m ceiling was moved onto this tower after it was built. 

Figure \ref{fig:instrument setup} (lower left) showed the Daocheng site's setup. The 10m automated weather station was bound to a previously built 22m tower. All other instruments were installed on the 7.5m concrete observation tower shown as the white building in right panel of \ref{fig:topo site}(b). 

Figure \ref{fig:instrument setup} (right) showed the layout of instruments at Muztagh Ata site. Because the 11m tower had the same design as Ali's 11m tower, it also suffered from the same structural instability problem. Two DIMMS (a NIAOT DIMM and a French DIMM) were setup on the ground (upper figure in figure \ref{fig:instrument setup} (right)). The French DIMM was later moved to the platform of the 11m tower while the NIAOT DIMM was moved to the platform of a 6m tower (details of these changes are described in detail in section \ref{sec:results-seeing}).

\section{Two years' results} \label{sec:results}
In this section, we will show top level monitoring results from data gathered between March 10th, 2017 and March 10, 2019. The selection of this period is to ensure that all equipment from all three sites were in operation mode for whole two years. Unless monitoring equipment was found faulty with significant system error, all data gathered in this period were treated as effective data. The number of days with effective data for different types of instruments during this 2-years period for all three sites are listed in table \ref{tab:dataavailability}. Data from all candidate sites were gathered and transmitted everyday by local teams to a remote central data server at National Astronomical Observatories, Chinese Academy of Sciences (NAOC), and was processed by the independent data analysis team using same scripts across all sites to ensure minimal possibility of bias. However even data processed in such a method, there were cases, such as DIMM, at certain sites that equipment was temporarily moved to an adjacent place for a brief period time. In this article, we will note what and when such cases happen. Besides, statistics are done in a monthly manner to separate and alleviate their impacts on the final results.

\begin{table}
\centering
\begin{tabular}{lccc}
\hline
Instruments & Ali & Daocheng & Muztagh Ata\\
\hline\hline
DIMM & 457 & 357 & 422\\
ASCA & 697 & 590 & 694 \\
Weather station & 675 & 559 & 705\\
SBM & 603 & 629 & 673\\
\hline
\end{tabular}
\caption{Data availability for instruments at the candidate sites between March 10th, 2017 and March 10, 2018.}
\label{tab:dataavailability}
\end{table}

\subsection{Seeing}\label{sec:results-seeing}
\begin{figure}
\begin{center}
\includegraphics[width=0.5\columnwidth]{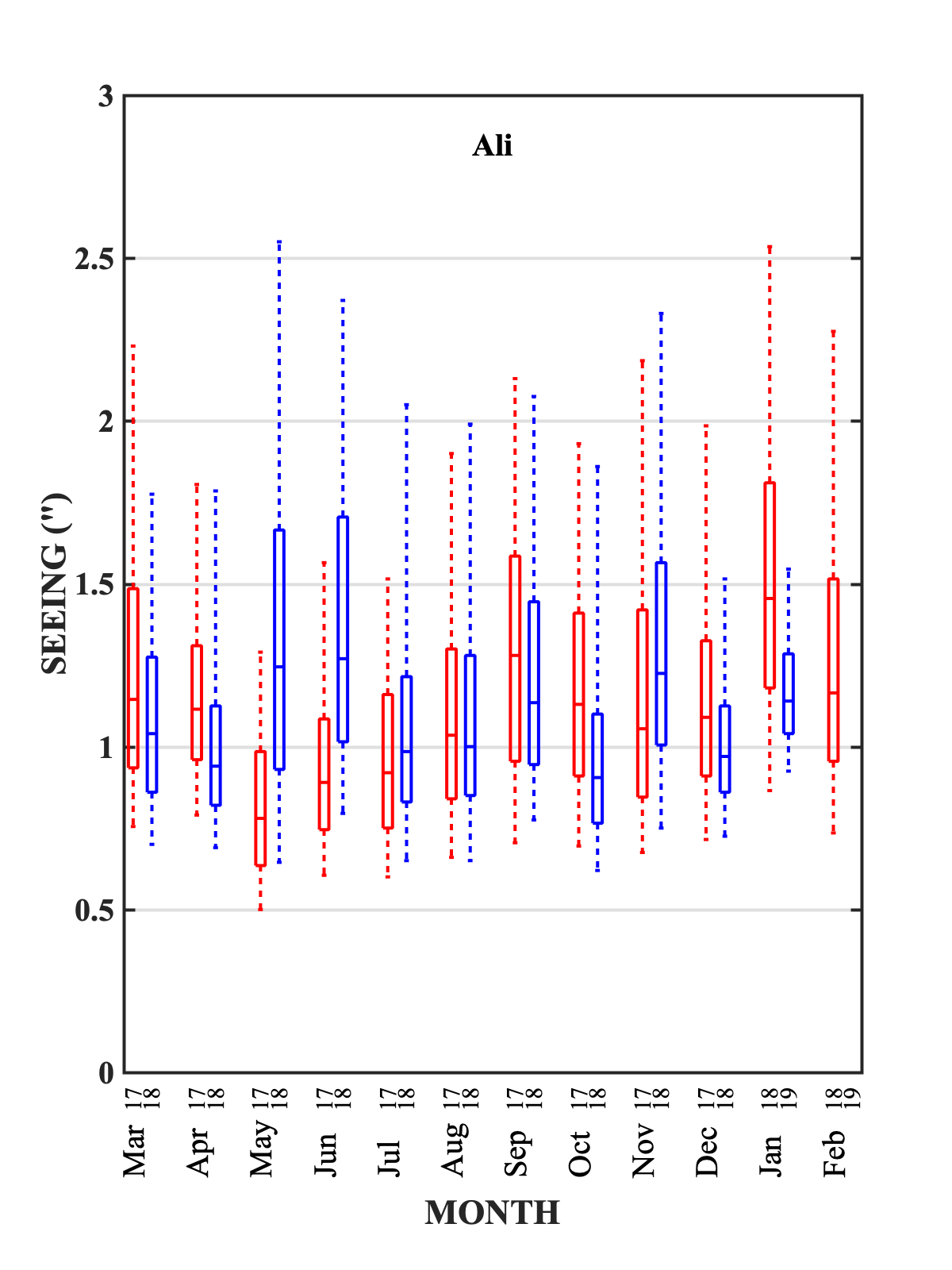}
\caption{Box plot of seeing measurements of Ali site. The upper tip, upper top of the box, mid-bar in the box, bottom of the box and lower tip represents 95\%, 75\%, 50\%, 25\%, and 5\% of the measured data for each month respectively. Red box represents data from March 2017 to February 2018, and blue box represents data from March 2018 to February 2019.}
\label{fig:Ali seeing}
\end{center}
\end{figure}

Ali's seeing was measured at 3 different heights: 3m roof of the observation building, 5m tower and 11m tower as shown in the previous section. Due to the reinforcement construction work of 11m tower, before May 2017, seeing measurements came from DIMM on the 3m roof top. After the 5m tower was constructed and before the 11m tower reinforcement work was done, seeing measurements mainly came from the DIMM on the 5m tower, but if any malfunction showed up on that DIMM, we used the 3m DIMM's measurements to fill that day's gap. After 11m tower was reinforced, data completely came from DIMM on the 11m tower. Cross comparison results of 3m and 5m measurements showed unnoticeable difference. Actually, even seeings measured at 5m tower and 11m tower showed minor difference (refer to \cite{Cao2019a}).  Figure \ref{fig:Ali seeing} shows the monthly statistics of these combined seeing measurements. 

\begin{figure}
\begin{center}
\includegraphics[width=0.5\columnwidth]{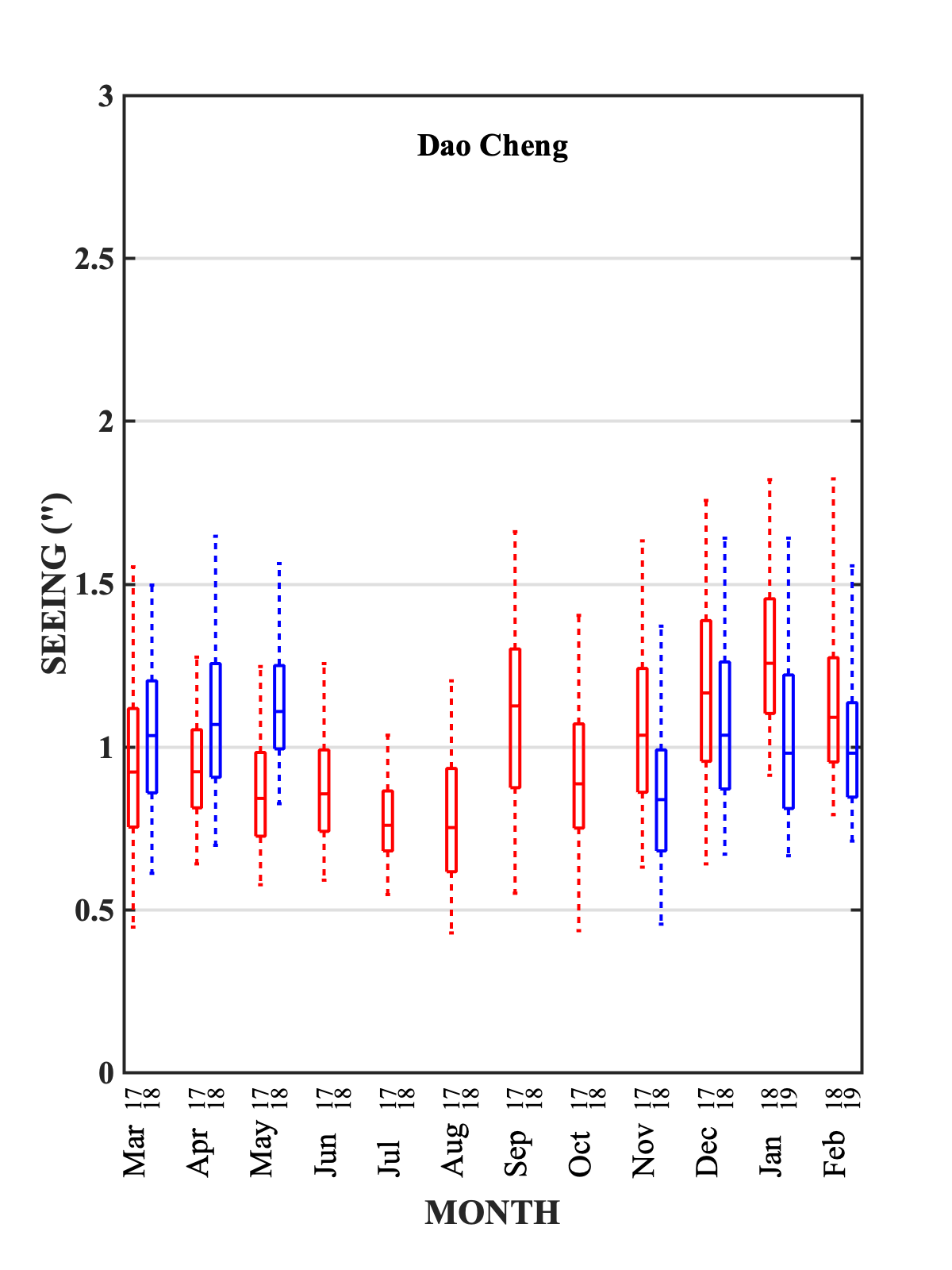}
\caption{Seeing measurements of Daocheng site. Representation method is the same as \ref{fig:Ali seeing}. DIMM was inoperable during June 2018$\sim$October 2018 at the site, which caused lack of data in this period.}
\label{fig:Daocheng seeing}
\end{center}
\end{figure}

Daocheng's seeing was measured at the 7.5m tower of the site. During July 2018 and October 2018, DIMMs at the site were out of operation due to system failure which caused the missing of data in that period. Figure \ref{fig:Daocheng seeing} shows the monthly statistical seasonal variation of the site's seeing.

\begin{figure}
\begin{center}
\includegraphics[width=0.5\columnwidth]{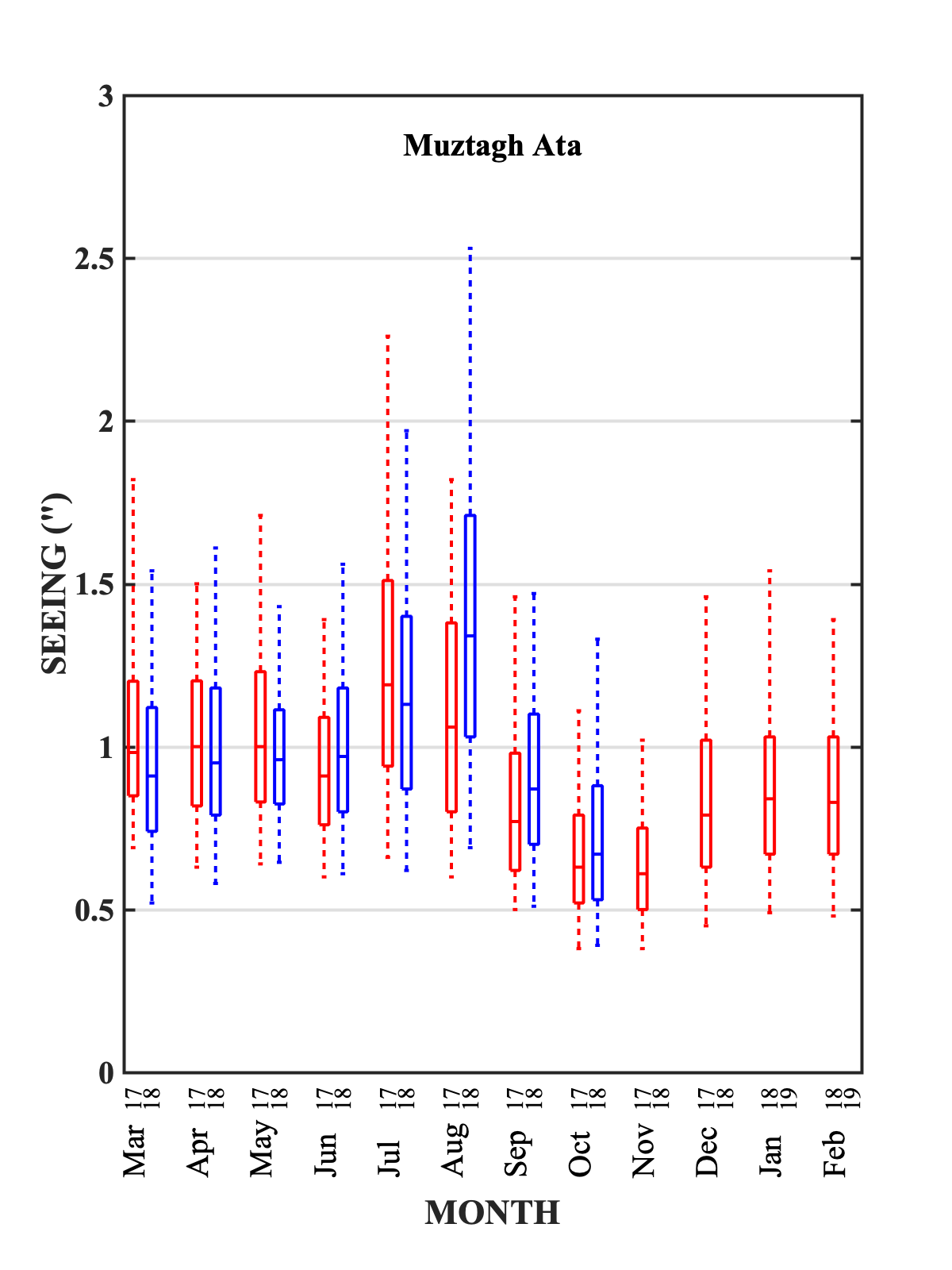}
\caption{Seeing measurements of Muztagh Ata site. From March 2017 to September 2017, measurements were done on the ground. In October 2017, simultaneous measurements of both at the ground and on the tower were conducted. It was found that measurements on the ground were 0.3" worse than on the tower at that time.}
\label{fig:Muztagata seeing}
\end{center}
\end{figure}

Measurements from Muztagh Ata is a bit complicated. NIAOT DIMM was installed in 12th March 2017 and kept running on ground until 15th November 2017. It was then moved to the top of the 6-meter tower. French DIMM was installed in 15th April 2017. It kept running on the ground until 23rd June 2017, and later was moved to the top of the 11-meter tower. It was soon found out that the structure strength of the tower was not sufficient for stable measurement of seeing. Therefore, the tower was reinforced with concrete the same as Ali by October, 2017. Simultaneous measurements on the ground and on the 11m tower in that duration showed that seeing measurements on the tower were approximately 0.3" better. After October 2017, all the measurements were done on the 11m tower. And then, after the rebuilding of 11-meter tower, for the purpose to ensure two DIMMs can be accomodated on the top of it simultaneously, NIAOT DIMM was installed on the 11-meter tower in 21st September 2018. The two DIMMs were then operated at 11 meters until 20th November 2018. Figure \ref{fig:Muztagata seeing} shows the monthly statistics of the site's seeing.

\begin{figure}
\begin{center}
\includegraphics[width=0.5\columnwidth]{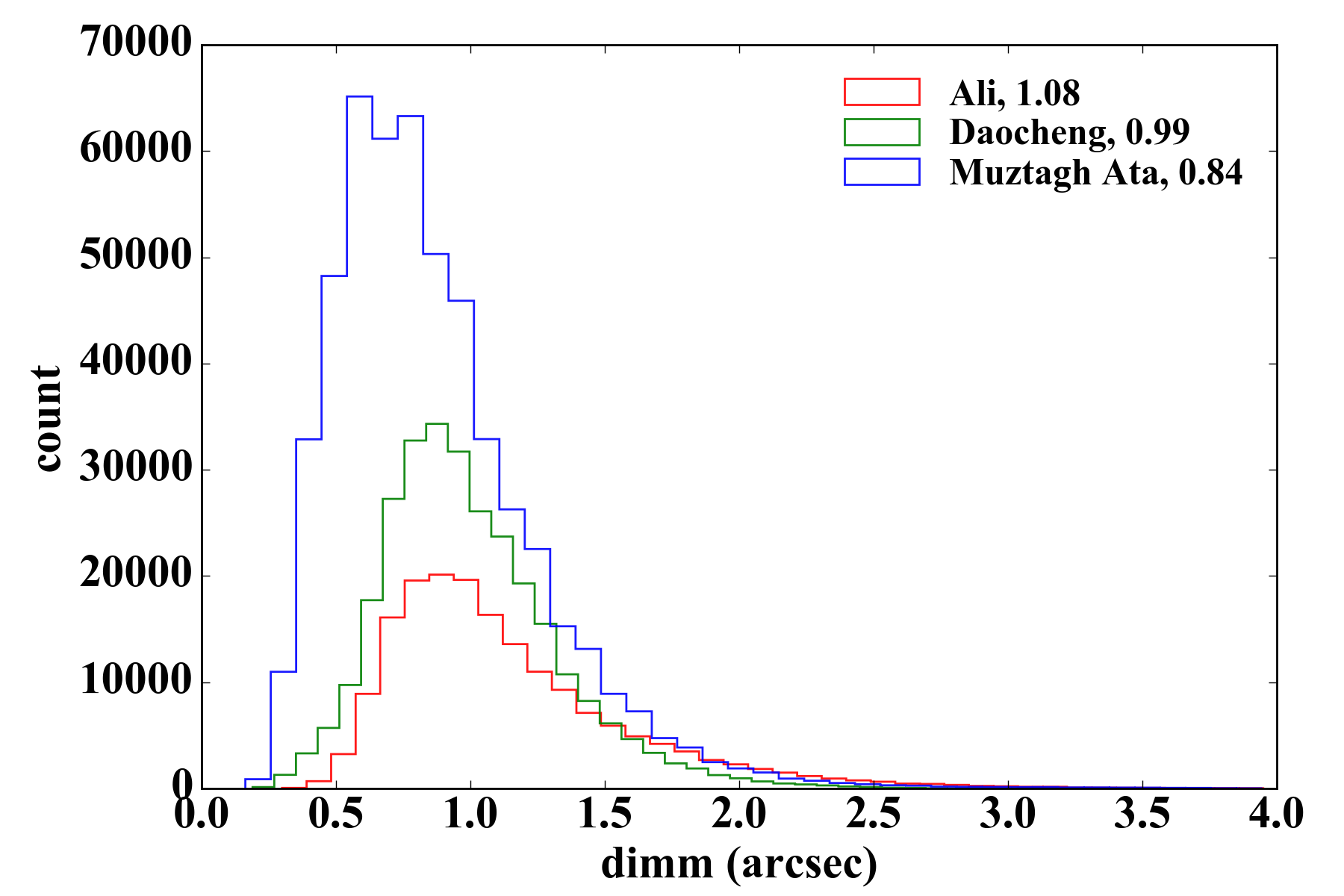}
\caption{Histogram of all seeing data for the three sites. Legends shows the median seeing for all three sites with all measurements.}
\label{fig:seeing histogram}
\end{center}
\end{figure}

Figure \ref{fig:seeing histogram} shows the histogram for all seeing measurements for all three sites. We would like to notify our reader that the lack of data in certain months (as can be seen from the other three monthly based box-plots) at certain sites could weigh the statistics into the direction of those months that have data, especially in regard of the median values shown in the legend of this figure. Therefore, in later section \ref{sec:discussion}, we chose to average the representative monthly median values for evaluating sites' performance.

\subsection{Cloudness}
\begin{figure}
\begin{center}
\includegraphics[width=0.5\columnwidth]{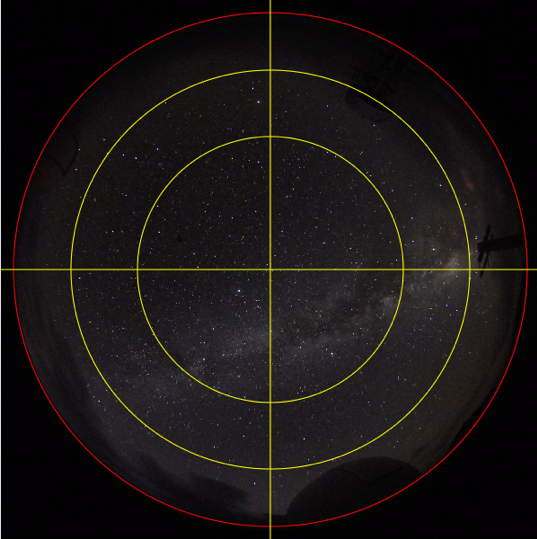}
\caption{An ASKA image to be processed. The two yellow inner circles have zenith angles of 44.7$^{\circ}$ and 65$^{\circ}$ respectively.  }
\label{fig:ASCA image}
\end{center}
\end{figure}

Cloudness was estimated from all the images taken by All Sky Camera manually. For everyday, from dusk to dawn, for every five minutes, a sky image pointing at Zenith direction was taken. The field of view of the image is 180$^{\circ}\times$180$^{\circ}$. Two circles are drawn within the image, with zenith angle of 44.7$^{\circ}$ and 65$^{\circ}$, as shown in figure \ref{fig:ASCA image}. Depending on how much clouds are within different circles and the quality of the image itself, five degrees of cloudness are defined:

\begin{itemize}
\item clear: no cloud within the inner circle and the outer circle;
\item outer: no cloud within the inner circle, has cloud within the outer circle 
\item inner: has cloud within both the inner and outer circles
\item covered: coverage of cloud within inner+outer circles are over 50\%
\item none: can not determine cloudness from image
\end{itemize}

The data analysis team adopted a similar method as TMT \cite{Skidmore2008}. Several people of the team watched movie generated with every frame of the ASCA images, judged and ranked the cloudness for each image manually. Statistics were done on these generated rankings for each site.    

We then defined an "observable night" if the night's cloudness was either clear, or outer, or inner for a continuous duration longer than 3 hours. A "good observable night" was defined if the cloudness of the night was either clear or outer for a continuous duration longer than 3 hours. 

\begin{figure}
\begin{center}
\includegraphics[width=0.5\columnwidth]{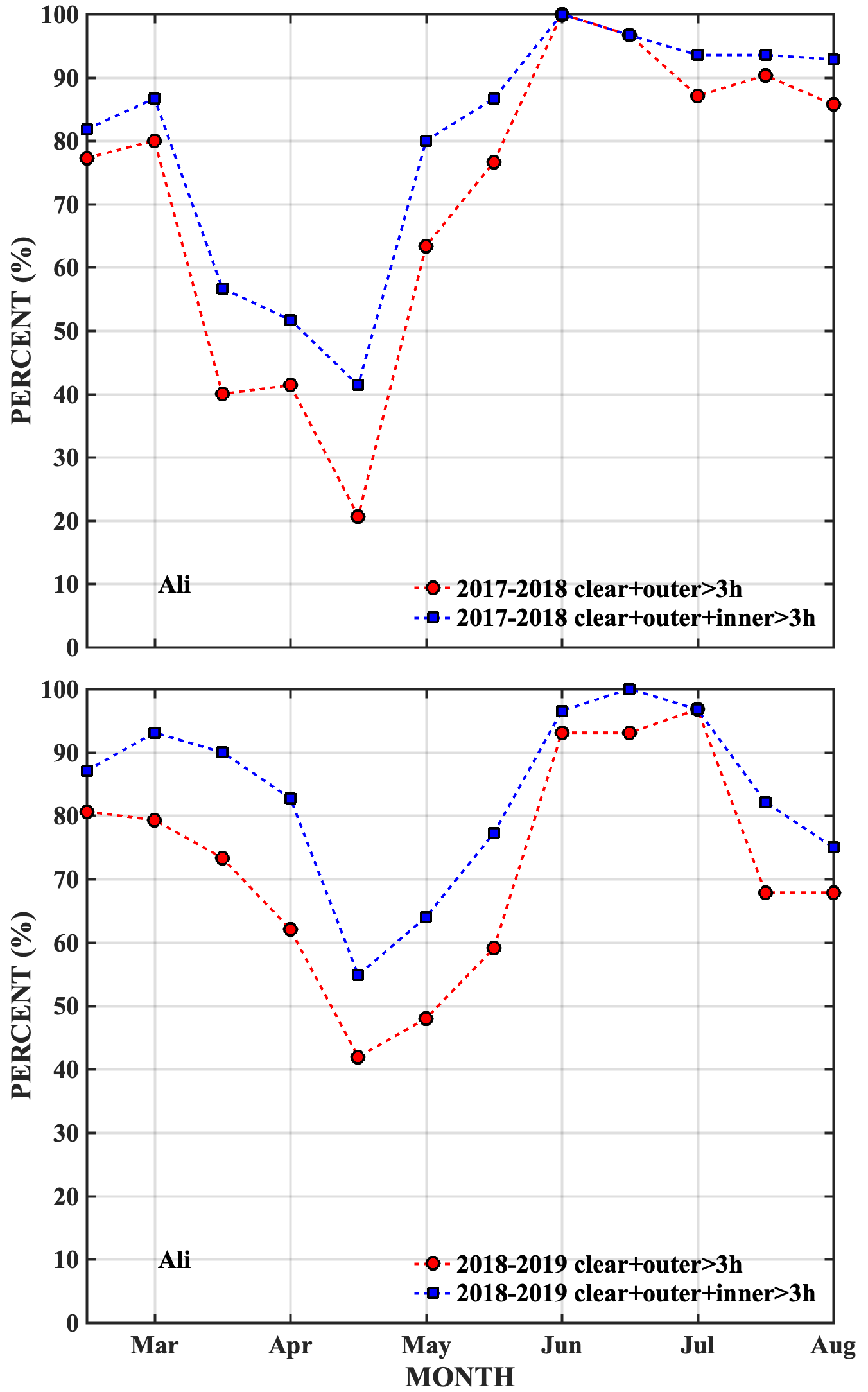}
\caption{Percentage of cloudness of Ali for March 2017$\sim$March 2018 (top) and March 2018$\sim$March 2019 (bottom). The blue dotted line shows the percentage of ``good observable nights" comparing to total number of nights with effective data in each month. The red dotted line shows the percentage of ``observable nights".}
\label{fig:cloud ali}
\end{center}
\end{figure}

\begin{figure}
\begin{center}
\includegraphics[width=0.5\columnwidth]{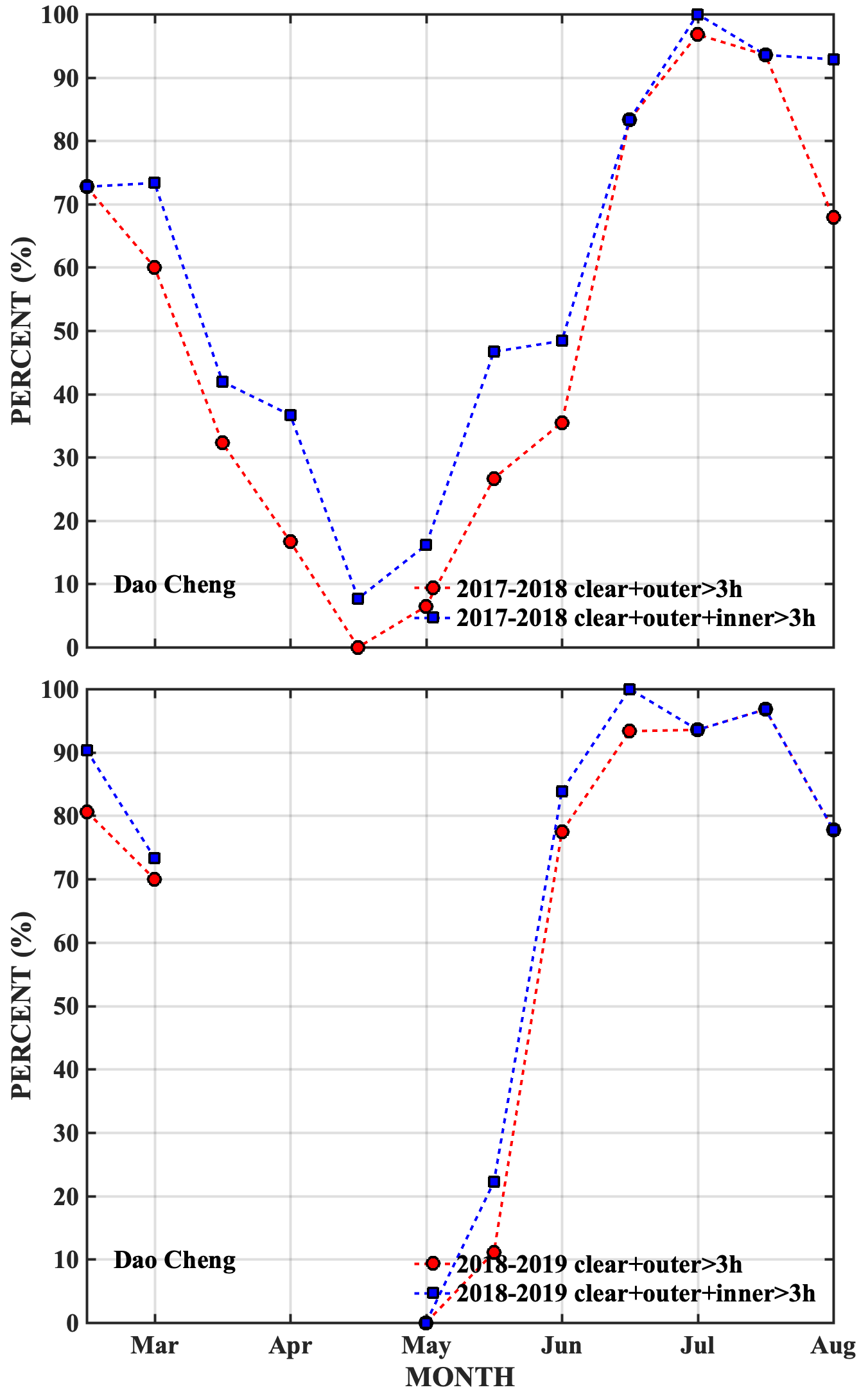}
\caption{Percentage of cloudness for Daocheng for March 2017$\sim$March 2018 (top) and March 2018$\sim$March 2019 (bottom). The blue dotted line shows the percentage of ``good observable nights" comparing to total number of nights with effective data in each month. The red dotted line shows the percentage of ``observable nights".}
\label{fig:cloud daocheng}
\end{center}
\end{figure}

\begin{figure}
\begin{center}
\includegraphics[width=0.5\columnwidth]{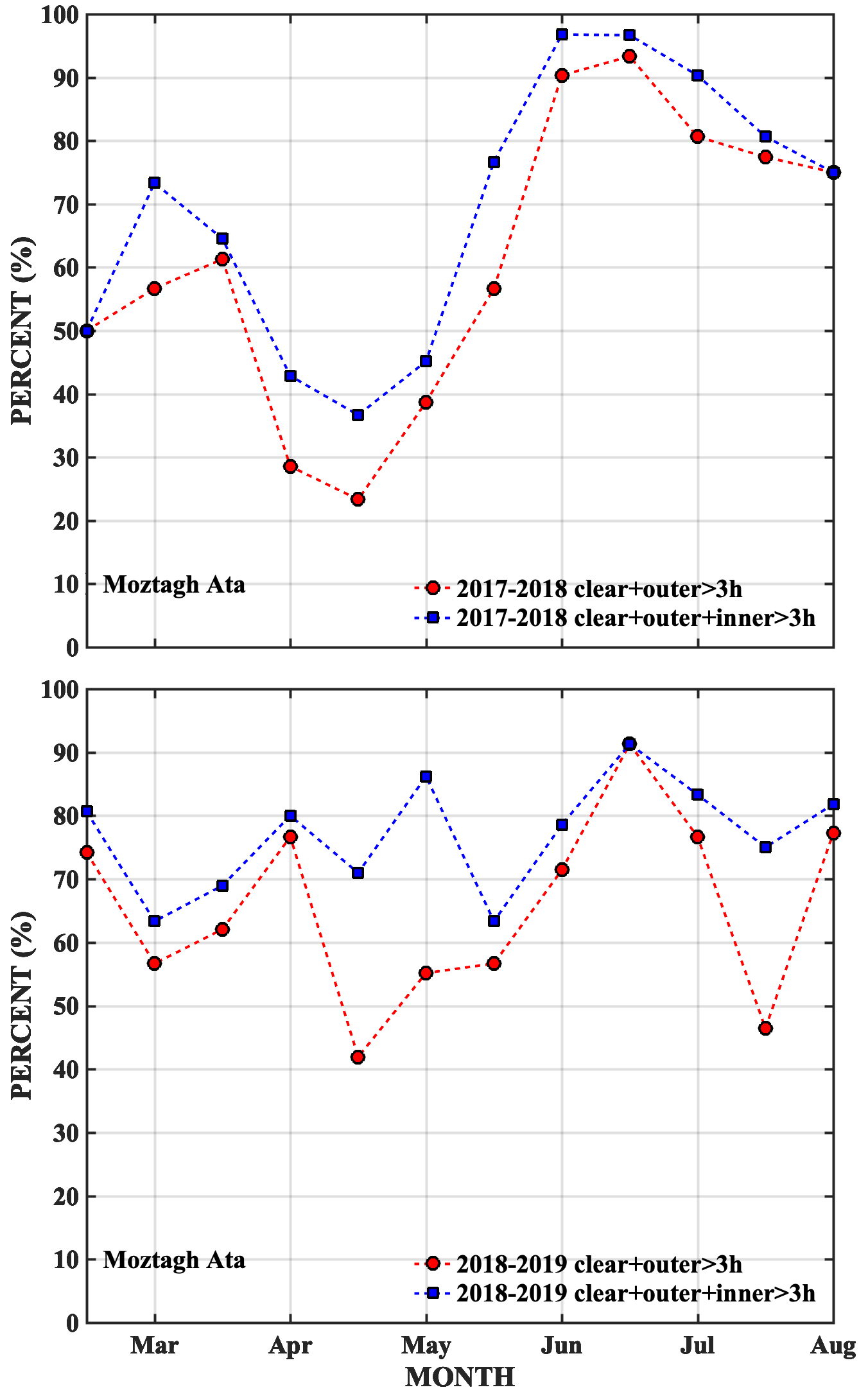}
\caption{Percentage of clear nights for Muztagh Ata for March 2017$\sim$March 2018 (top) and March 2018$\sim$March 2019 (bottom). The blue dotted line shows the percentage of ``good observable nights" comparing to total number of nights with effective data in each month. The red dotted line shows the percentage of ``observable nights".}
\label{fig:cloud muztagata}
\end{center}
\end{figure}

With data from March 10, 2017 to March 10, 2019, the number of these different qualities of nights and the total number of nights that had determinable ASCA images are listed in table \ref{tab:cloud}. In figure \ref{fig:cloud ali}, \ref{fig:cloud daocheng} and \ref{fig:cloud muztagata}, we shows the monthly variation of good night at Ali, Daocheng and Muztagata respectively. For Daocheng, during May 2018$\sim$July 2018, ASKA at the site was inoperable and data was missing in this period.

\begin{table}
\centering
\begin{tabular}{lccc}
\hline
Nights & Ali & Daocheng & Muztagh Ata\\
\hline\hline
Observable night & 568 & 401 & 507\\
Good observable night & 500 & 354 & 438 \\
Nights with effective data & 693 & 590 & 694 \\
\hline
\end{tabular}
\caption{Data availability for instruments at the candidate sites between March 10th, 2017 and March 10, 2019.}
\label{tab:cloud}
\end{table}

\subsection{Temperature, atmospheric pressure and relative humidity}
Temperature, relative humidity and atmospheric pressure were constantly measured by the automated weather station. The model of the weather stations used at candidate sites were the same. However, the measurement frequencies of these weather stations were different. Temperatures of Ali and Muztagh Ata were measured once every minute, Daocheng's temperature was measured once every 10 minutes. Another thing worth to mention when interpreting the plot of the data is that between April 2018$\sim$June 2018, Daocheng's weather station was inoperable. Therefore, in these three months Daocheng's data was missing. 

\begin{figure}
\begin{center}
\includegraphics[width=0.5\columnwidth]{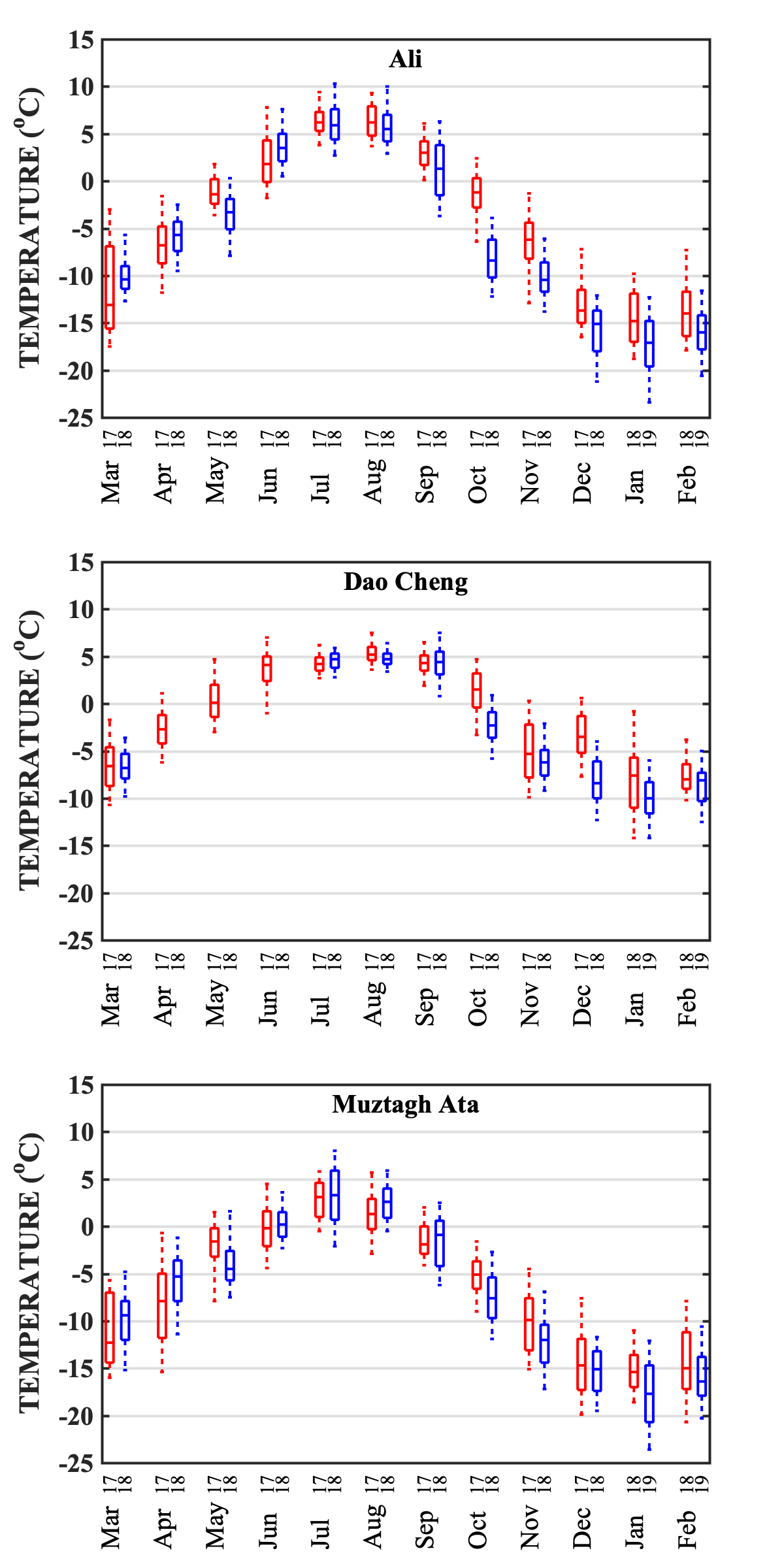}
\caption{Night time temperature monthly statistics. (Top to bottom) Ali, Daocheng, Muztagh Ata. The upper tip, upper top of the box, mid-bar in the box, bottom of the box and lower tip represents 95\%, 75\%, 50\%, 25\%, and 5\% of the measured data for each month respectively. Red box represents data from March 2017 to February 2018, and blue box represents data from March 2018 to February 2019.}
\label{fig:temperature night time}
\end{center}
\end{figure}

In figure \ref{fig:temperature night time}, we showed the monthly variations of night temperatures at Ali, Daocheng and Muztagh Ata. 

\begin{figure}
\begin{center}
\includegraphics[width=0.5\columnwidth]{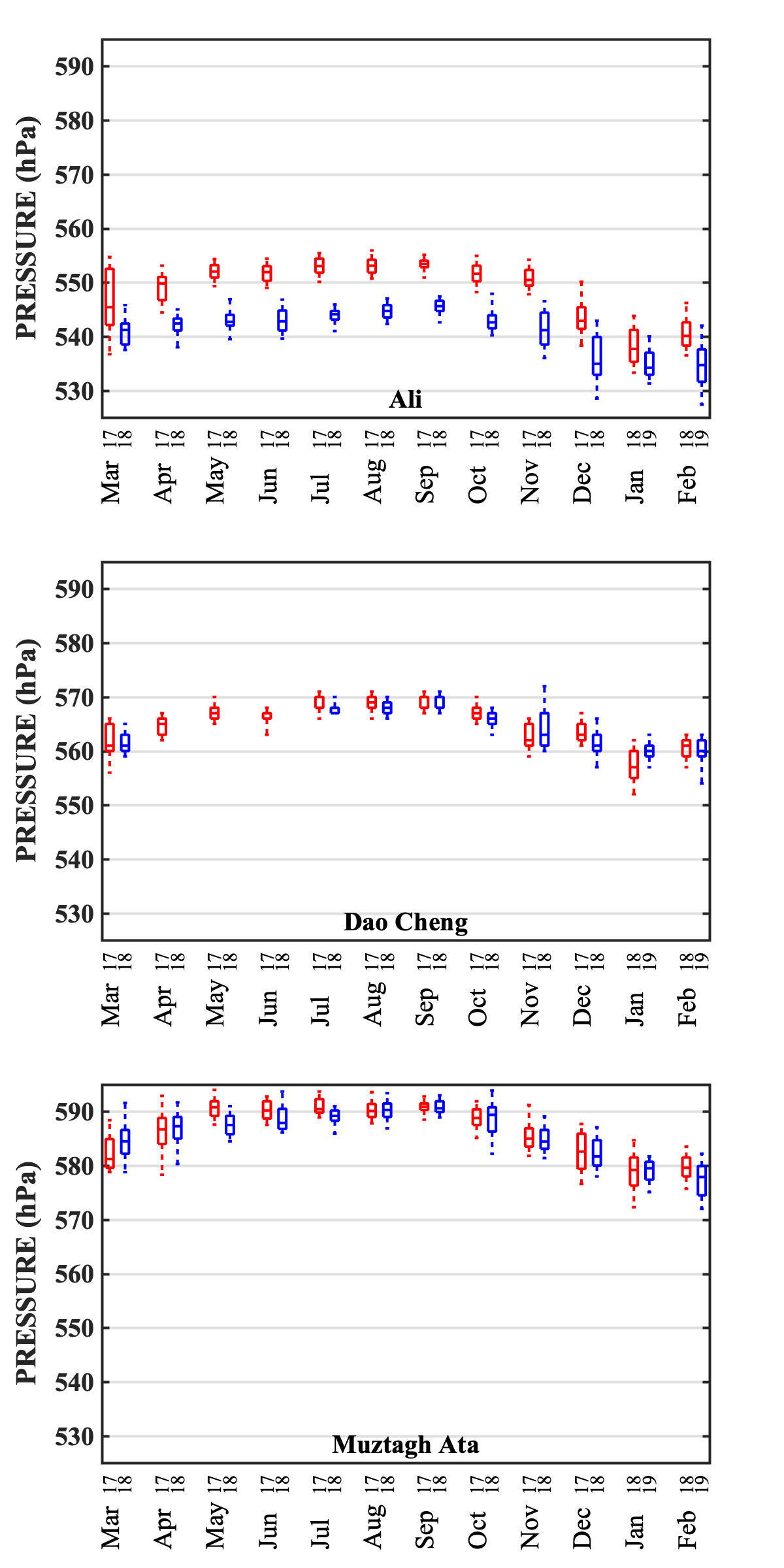}
\caption{Night time pressure statistics. (Top to bottom) Ali, Daocheng, Muztagh Ata. The upper tip, upper top of the box, mid-bar in the box, bottom of the box and lower tip represents 95\%, 75\%, 50\%, 25\%, and 5\% of the measured data for each month respectively. Red box represents data from March 2017 to February 2018, and blue box represents data from March 2018 to February 2019.}
\label{fig:pressure night time}
\end{center}
\end{figure}

Monthly statistics of night time atmospheric pressure measurements for the three sites were shown in  figure \ref{fig:pressure night time}.

\begin{figure}
\begin{center}
\includegraphics[width=0.5\columnwidth]{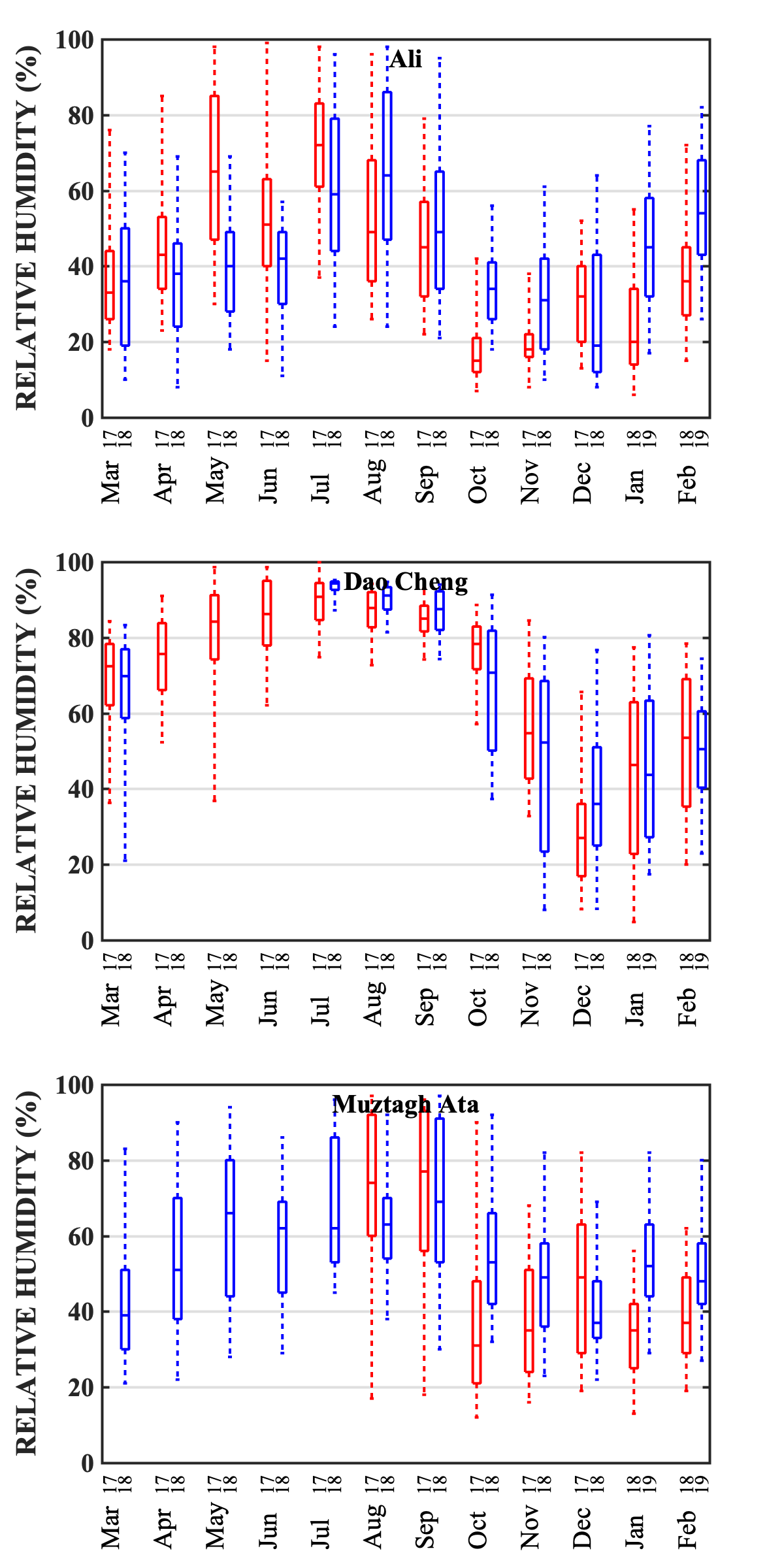}
\caption{Night time humidity statistics. (Top to bottom) Ali, Daocheng, Muztagh Ata. The upper tip, upper top of the box, mid-bar in the box, bottom of the box and lower tip represents 95\%, 75\%, 50\%, 25\%, and 5\% of the measured data for each month respectively. Red box represents data from March 2017 to February 2018, and blue box represents data from March 2018 to February 2019.}
\label{fig:humidity night time}
\end{center}
\end{figure}

Figure \ref{fig:humidity night time} shows the relative humidity measurements for night time. The humidity sensor at Muztagh Ata was in malfunction until August 3rd, 2017. Therefore, for Muztagh Ata, relative humidity data after August 3rd, 2017 were adopted for statistics.

\subsection{Wind speed and direction}
\begin{figure}
\begin{center}
\includegraphics[width=0.5\columnwidth]{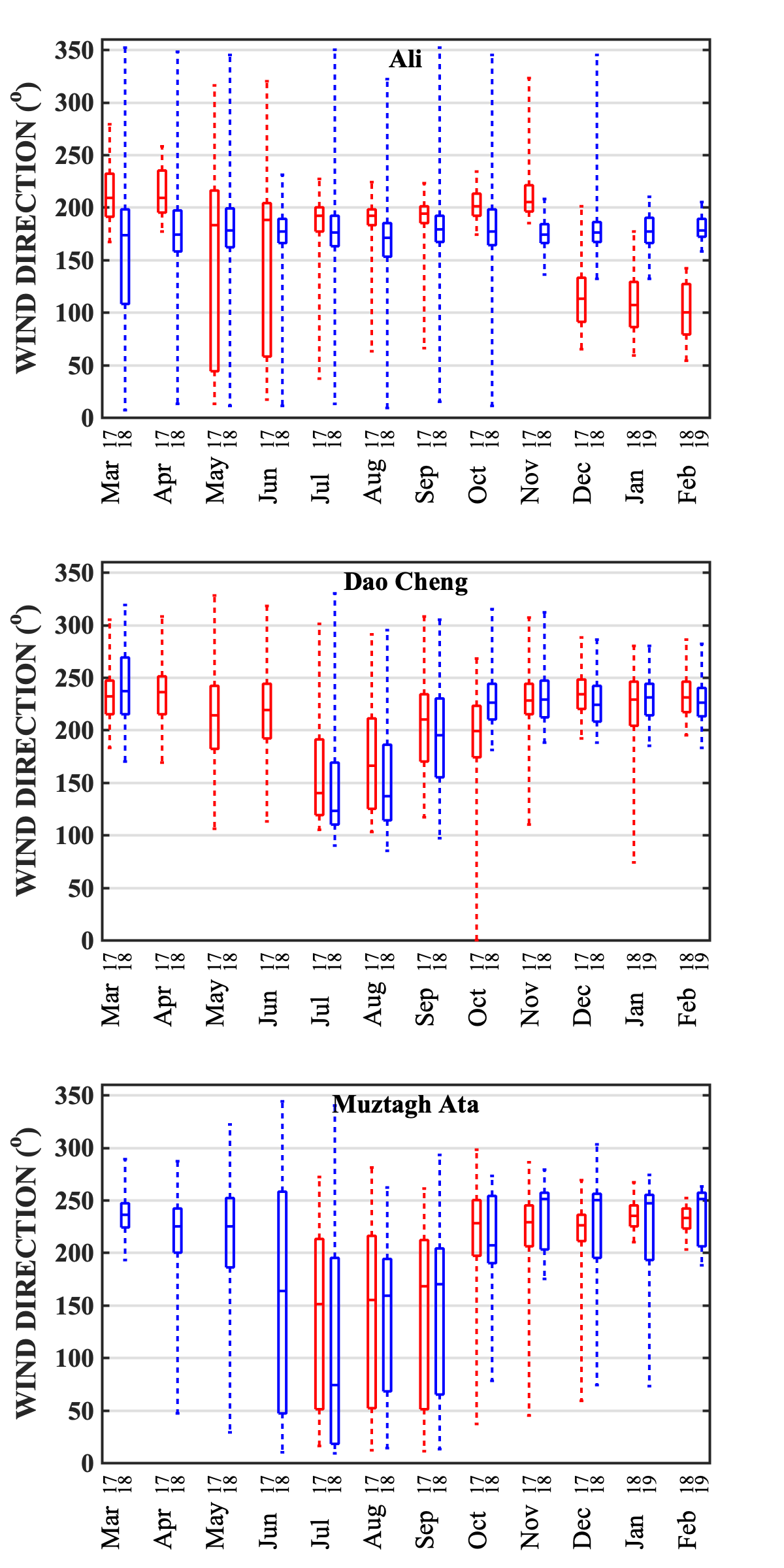}
\caption{Monthly variation of wind direction ($0^o\sim360^o$) at the site. (Top to bottom) Ali, Daocheng and Muztagh Ata. The upper tip, upper top of the box, mid-bar in the box, bottom of the box and lower tip represents 95\%, 75\%, 50\%, 25\%, and 5\% of the measured data for each month respectively. Red box represents data from March 2017 to February 2018, and blue box represents data from March 2018 to February 2019.}
\label{fig:wind direction at night}
\end{center}
\end{figure}

\begin{figure}
\begin{center}
\includegraphics[width=0.5\columnwidth]{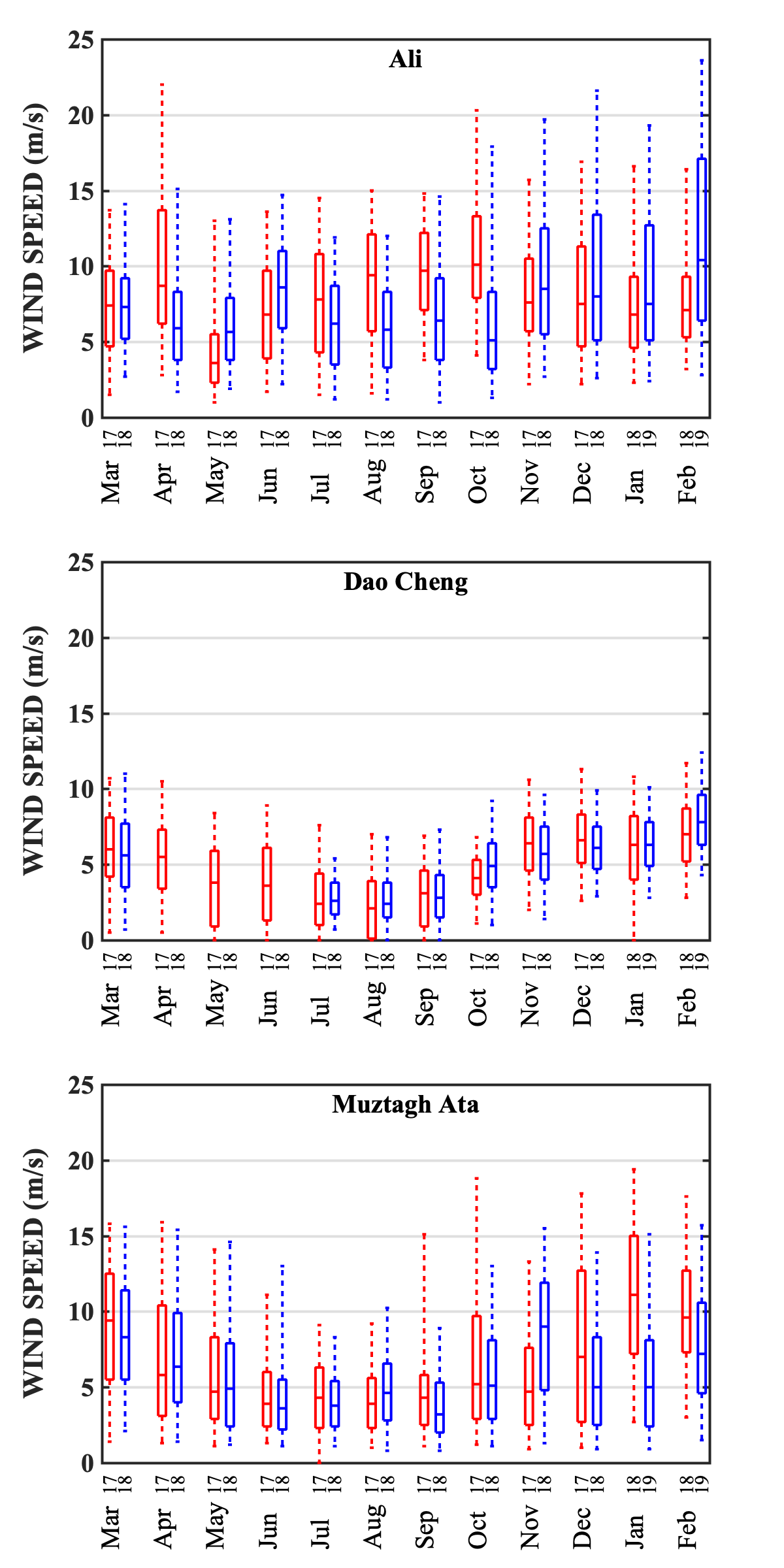}
\caption{Monthly variation of wind speed at the site. (Top to bottom) Ali, Daocheng, Muztagh Ata. The upper tip, upper top of the box, mid-bar in the box, bottom of the box and lower tip represents 95\%, 75\%, 50\%, 25\%, and 5\% of the measured data for each month respectively. Red box represents data from March 2017 to February 2018, and blue box represents data from March 2018 to February 2019.}
\label{fig:wind speed at night}
\end{center}
\end{figure}

Wind speed and direction were measured with anemometers mounted on top of automated weather stations' 10m masts for every minute. Figure \ref{fig:wind direction at night} and figure \ref{fig:wind speed at night} shows the monthly statistics of wind speed and wind direction at night for each of the three sites.

\subsection{Sky background}
\begin{figure}
\begin{center}
\includegraphics[width=0.5\columnwidth]{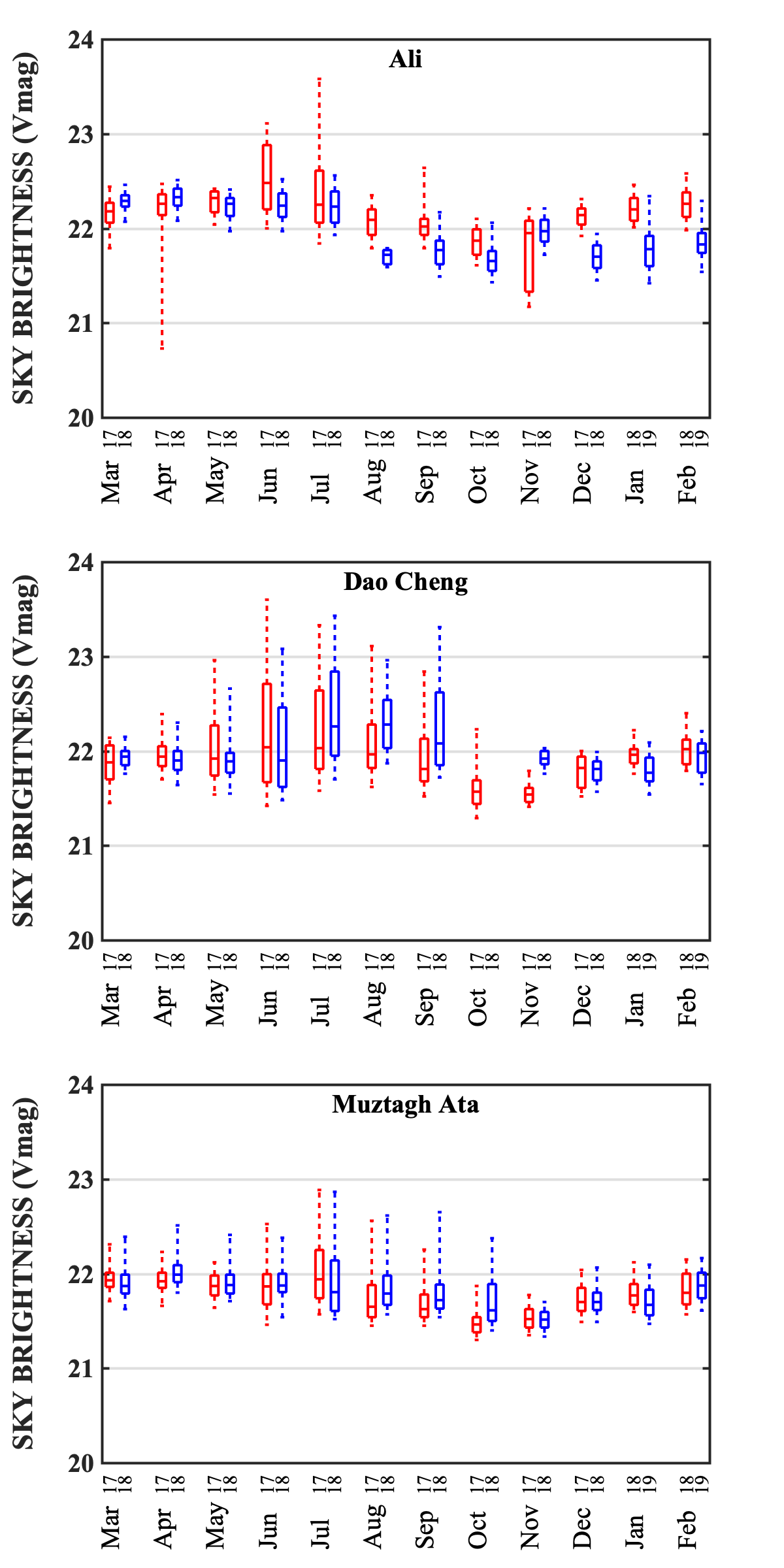}
\caption{Sky background in V magnitude for no moon nights. (Top to bottom) Ali, Daocheng, Muztagh Ata. The upper tip, upper top of the box, mid-bar in the box, bottom of the box and lower tip represents 95\%, 75\%, 50\%, 25\%, and 5\% of the measured data for each month respectively. Red box represents data from March 2017 to February 2018, and blue box represents data from March 2018 to February 2019.}
\label{fig:background}
\end{center}
\end{figure}

V band sky background were continuously measured by the aforementioned SBM. Considering lights from the moon have impacts on the brightness of night sky, here in this paper we only reported sky background statistics from no moon nights. Figure \ref{fig:background} shows the V band sky background of Ali, Daocheng and Muztagh Ata for those nights. 

\subsection{Precipitable water vapor}
Precipitable Water Vapor (PWV) values were measured with a LHATPRO. Until now, only one such instrument was acquired, and it was equipped much later than all other equipments. Therefore, PWV values for each site were measured with limited number of nights at the moment. Because local cloud can have a strong influence on PWV measurements, in this paper, we only show results obtained in clear nights.

\begin{figure}
\begin{center}
\includegraphics[width=0.5\columnwidth]{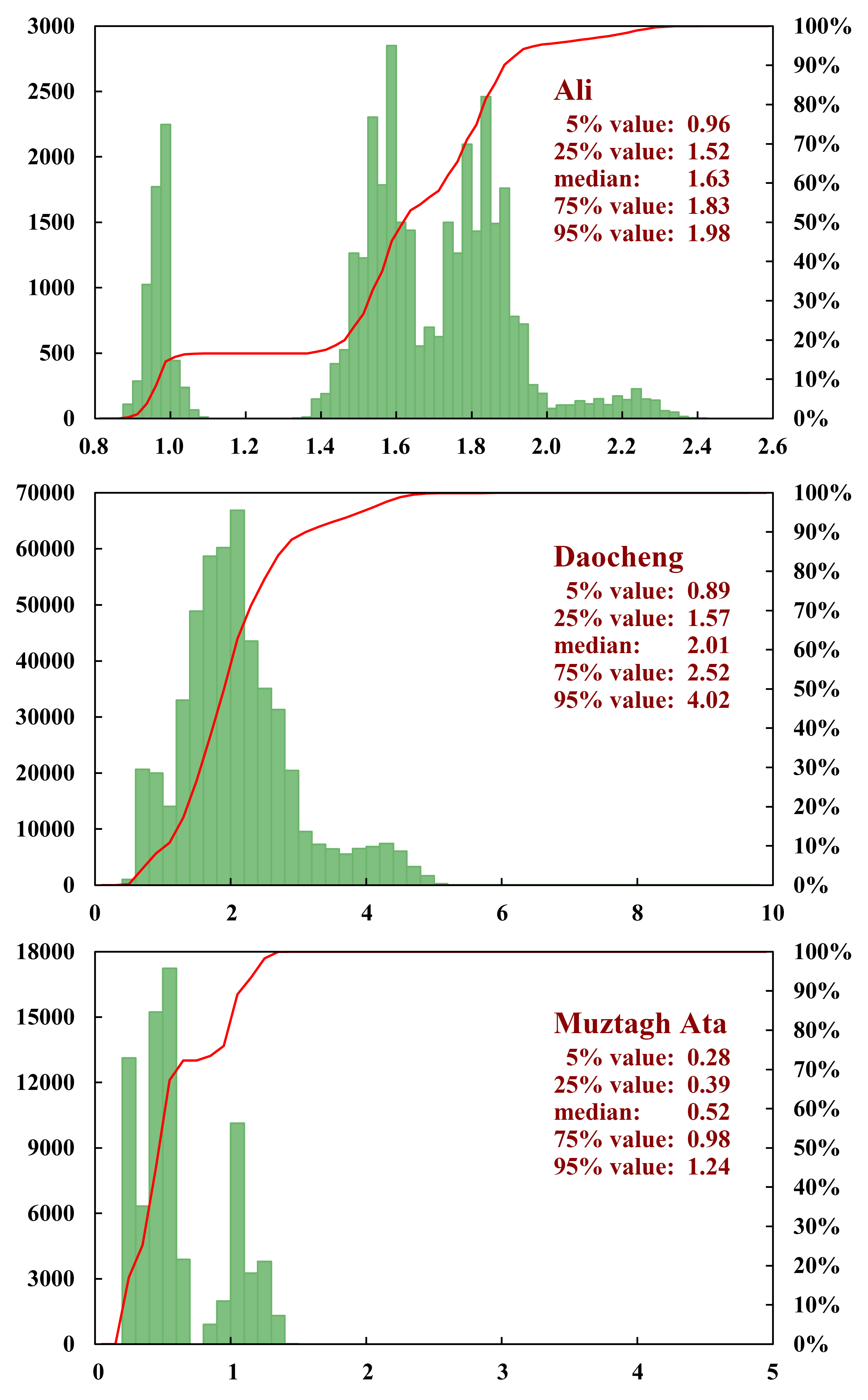}
\caption{Precipitable water vapor in clear nights for Ali, Daocheng and Muztagh Ata from top to bottom.}
\label{fig:pwv}
\end{center}
\end{figure}

For Ali, PWV monitoring was conducted in a very short time (March 19th, 2018 to March 22nd, 2018). Data obtained was significantly less than the other two sites. The PWV statistics for this site is shown here just for completeness, but we would like to remind our readers that this result may not be representative for the site. Because of this, we did not list any Ali's PWV value in table \ref{tab:result summary}.  
For Daocheng, all measurements were done between February 10th, 2018 and March 5th, 2018. Its statistics is shown in figure \ref{fig:pwv} mid panel.
For Muztagh Ata, measurements were conducted between January 22nd, 2018 and February 4th, 2018. The bottom panel of figure \ref{fig:pwv} shows its statistics. 

\section{Discussion} \label{sec:discussion} 

Table \ref{tab:result summary} is a summary of monitored performances for all three candidate sites with two year's data between March 10th, 2017 and March 10th, 2019. From these annual average results, one could see that Ali had the highest good observable nights fraction and darkest night sky than the other two sites. All three sites showed impacts of rain seasons with different degree. The most interesting part is Muztagh Ata, where although in the first year, cloud coverage was higher in summer, in the second year, the impact of rain season was lessened. We refer all our readers to \cite{Cao2019b} which might show some insights to this behavior. As regarding to seeing, the annual average values for all three sites had reached $\sim$1 arcsec, while Muztagh Ata's seeing was the best among the three, especially considering that the site's best seeing season was coinciding with cloudless time of the year (October to December). Daocheng had the highest night temperature of -1.7 Celsius, and the other two sites were similar at around -6 Celsius. For relative humidity, Daocheng was 70.15\% and was much higher than the other two sites. Ali had the highest annual averaged wind speed of 7.4m/s than the other two sites which were around 5m/s.  With current relatively short PWV measurements, Muztagh Ata had the lowest precipitable water vapors of 0.52mm among all three sites in roughly the same season.

\begin{table}
\centering
\begin{tabular}{lccc}
\hline
Parameters & Ali & Daocheng & Muztagh Ata\\
\hline\hline
Cloudness: \\
  good observable $>$ 3hr (\%) & 71.76 & 51.72 & 63.27 \\
  observable $>$ 3hr (\%) & 81.71 & 58.89 & 73.00 \\
V band sky background (magV) & 22.07 & 21.91 & 21.76\\
Seeing (") & 1.17 & 1.01 & 0.90\\ 
Night temperature ($^\circ$) & -5.18 & -1.70 & -7.08 \\
Night Atm. pressure (hPa) & 540.92 & 564.54 & 586.08 \\
Night relative humidity (\%) & 41.25 & 70.15 & 52.88\\
Night wind speed (m/s) & 7.41 & 4.75 & 5.83 \\
Night wind direction & SW & SW & SW                                                                                                                                                                                                                                                                               \\
PWV (mm) & $-$ & 2.01 & 0.52 \\
\hline
\end{tabular}
\caption{Annual average with data from March 10th, 2017 to March 10th, 2018 for all three sites. The annual average is calculated by two steps: 1. Calculate the average value of each month with the two years monthly median values. If for some reason one year's median value is missing, the average value is the other year's median value. Thus, we could get a monthly average value for each month for these two years. 2. Calculate the average value of the monthly average values calculated in the last step to represent the annual average value. }
\label{tab:result summary}
\end{table}

It is still too early to jump to a conclusion that is scientifically sound and reasonable with only two years' data. It is also because of this reason, in this paper we are hesitant to directly combine the two years data all togather and do the statistics of it, but rather showing both years' weather variations. For questions such as, why there is different behavior of cloud at Muztagh Ata in the second year from the first year, what would be the PWV like for all season at all three sites, they yearns more data for clear understanding. Two years data is not sufficient to answer whether one of the year or even both years were abnormal years? With current equipments we could also not give definitive answer to questions like, how are the performances of these sites in infrared, or is there a possibility that a much better site exists near current ones? All these questions alike could only be answered by actual measurements from more monitoring instruments that will be installed soon and with more monitoring data of longer period. 

\section*{Acknowledgements}

Site testing is, to say, not the only but still among one of the toughest works in the business, especially considering this was the first time in China that several teams from different institutes were being coordinated to do all their best to simultaneously monitoring all three sites in a systematic way. Many colleagues and friends had generously offered their helps and we hereby thank everyone involved. We would also like to thank Dr. Marc Sarazin from European Southern Observatory, and Prof. Paul Hickson from University of British Columbia for their guidances and great helps. 

The research is partly supported by the Operation, Maintenance and Upgrading Fund for Astronomical Telescopes  and  Facility  Instruments,  budgeted  from  the  Ministry  of Finance of China (MOF) and administrated by the Chinese Academy of Sciences (CAS). The research is also supported by National Natural Science Foundation of China No.11873081. 



\label{lastpage}


\begin{thebibliography}{99}

\bibitem[ALCOR-SYSTEM 2016]{alcor} ALCOR-SYSTEM, 'DIMM seeing monitor', 2016, \url{http://www.alcor-system.com/new/SeeingMon/DIMM_Complete.html} 

\bibitem[Cao et al. 2019a]{Cao2019a} Cao, Z. H., Li, J., Zhao, Y. H. et al., 2019, \raa, this issue

\bibitem[Cao et al. 2019b]{Cao2019b} Cao, Z. H., Liu, L. Y., Yao, Y. Q. et al., 2019, \raa, this issue	

\bibitem[Cui et al. 2018]{Cui2018} Cui, X. Q., Zhu, Y. T., Liang, M. et al., 2018, in Proc. of SPIE, Vol. 10700, Ground-based and Airborne Telescope VII, ed., Jason S. (SPIE), 107001P

\bibitem[Huayun 2016]{huayun} China Huayun, 'Huayun CAWS 600 automatic weather station', 2016, \url{http://www.cnhyc.com/showcpzs.asp?id=489}

\bibitem[Liu et al. 2012]{Liu2012} Liu, L. Y., Yao, Y. Q., Vernin, J. et al., 2012, in Proc. of SPIE, Vol. 8444, Ground-based and Airborne Telescope IV, ed., Spyromilio J., (SPIE), 844464

\bibitem[Liu et al. 2015]{Liu2015} Liu, L. Y., Yao, Y. Q., Vernin, J. et al., 2015, in JPCS, Vol. 595, 012019
	
\bibitem[Liu et al. 2016]{Liu2016} Liu, Y., Song, T. F., Zhang, X. F. et al., 2016, in Proc. of IAU, Vol. 320, Solar and Stellar Flares and their Effects on Planets, 447

\bibitem[Liu et al. 2019]{Liu2019} Liu, L. Y., Yao, Y. Q., Yin, J. et al., 2019, \raa, this issue
	
\bibitem[Qian et al. 2015]{Qian2015} Qian, X., Yao, Y. Q., Wang, H. S. et al., 2015, PKAS, 30, 695
	
\bibitem[Radiometer Physics 2014]{rpg} Radiometer Physics, 'Humidity and Temperature Profilers', 2014, \url{http://alturl.com/3vvyw}
	
\bibitem[Sarazin et al. 1990]{Sarazin1990} Sarazin, M., Roddier, F., 1990, \aap, Vol. 227, 294

\bibitem[Skidmore et al. 2008]{Skidmore2008} Skidmore, W., Sch{\"o}ck, M., Magnier, E. et al, 2008, in Proc. of SPIE, Vol. 7012, 701224
	
\bibitem[Song et al. 2019]{Song2019} Song, T. F., Liu, Y., Wang, J. X. et al, 2019, \raa, this issue

\bibitem[Su et al. 2016]{Su2016} Su, D. Q., Bai, H., Liang, M. et al., 2016, MNRAS, 460, 2286
	
\bibitem[Unihedron 2016]{Unihedron2016} Unihedron, 'Sky quality meter - LE', 2016, \url{http://unihedron.com/projects/sqm-le/}
	
\bibitem[Wang et al. 2015]{Wang2015} Wang, H. S., Yao, Y. Q., Liu, L. Y. et al., 2015, in JPCS, Vol. 595, 012037	
	
\bibitem[Wang et al. 2019]{Wang2019} Wang, J. F., Tian, J. F., Li, T. R. et al., 2019, \raa, this issue

\bibitem[Wu et al. 2016]{Wu2016} Wu, N., Liu, Y., \& Zhang, H. M., 2016, AcASn, 57, 729

\bibitem[Xu et al. 2019a]{Xu2019a} Xu, J., Esamdin, A., Pu, G. X. et al, 2019, \raa, this issue
	
\bibitem[Xu et al. 2019b]{Xu2019b} Xu, J., Esamdin, A., Pu, G. X. et al, 2019, \raa, this issue
	
\bibitem[Yao et al. 2012]{Yao2012} Yao, Y. Q, Wang, H. S., Liu, L. Y. et al., 2012, in Proc. of SPIE, Vol. 8444, 844413
	
\bibitem[Yao et al. 2013]{Yao2013} Yao, Y. Q, Wang, Y. P., Liu, L. Y. et al., 2013, in NARIT Conf. Ser., Vol. 1, the 11th Asian-Pacific Regional IAU Meeting 2011, ed., Komonjinda S. (IAU), 1

\bibitem[Yao et al. 2014a]{Yao2014} Yao, Y. Q., Zhou, Y. H., Liu, L. Y. et al., 2014, in JPCS, Vol. 595, 

\bibitem[Yao et al. 2014b]{Yao2015} Yao, Y. Q., Zhou, Y. H., Liu, L. Y. et al., 2014, in JPCS, Vol. 595, 012038

\bibitem[Yin et al. 2014]{Yin2015} Yin, J., Yao, Y. Q., Liu, L. Y. et al., 2014, in JPCS, Vol. 595, 012040	

\end{thebibliography}
\end{document}